\documentclass{article}

\usepackage[utf8]{inputenc}
\usepackage[T1]{fontenc}

\PassOptionsToPackage{table, dvipsnames}{xcolor}
\usepackage{xcolor}

\PassOptionsToPackage{colorlinks=true, citecolor=blue, linkcolor=blue, urlcolor=black}{hyperref}

\usepackage{arxiv}
\usepackage[utf8]{inputenc}
\usepackage[T1]{fontenc}
\usepackage{url}
\usepackage{booktabs}
\usepackage{nicefrac}
\usepackage{microtype}
\usepackage{lipsum}
\usepackage{graphicx}
\usepackage{natbib}
\usepackage{doi}
\usepackage{algorithm}
\usepackage{algpseudocode}%
\usepackage{listings}%
\usepackage{multirow}
\usepackage{multicol}
\usepackage{amsmath,amssymb,amsfonts}
\usepackage{amsthm}
\usepackage{mathrsfs}
\usepackage{float}
\usepackage{bm}
\usepackage{caption}
\usepackage{hyperref}

\usepackage{tabularx}
\usepackage{array}
\usepackage{pdflscape}

\title{Shallow-to-deep velocity model building via diffusion models-Part II: Realistic scenarios}

\author{
	\href{https://orcid.org/0000-0001-8868-7967}{\includegraphics[scale=0.06]{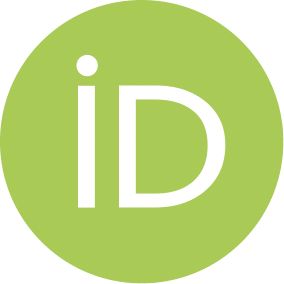}\hspace{1mm}Shijun~Cheng} \\
	Division of Physical Science and Engineering\\
	King Abdullah University of Science and Technology\\
	Thuwal 23955-6900, Saudi Arabia \\
        \And
    {Randy Harsuko} \\
	Division of Physical Science and Engineering\\
	King Abdullah University of Science and Technology\\
	Thuwal 23955-6900, Saudi Arabia \\
        \And
    {Tariq Alkhalifah} \\
	Division of Physical Science and Engineering\\
	King Abdullah University of Science and Technology\\
	Thuwal 23955-6900, Saudi Arabia \\
    [3ex]
  $^{*}$Corresponding author: \textbf{Shijun Cheng}~(\texttt{sjcheng.academic@gmail.com})
}

\renewcommand{\shorttitle}{Shallow-to-deep VMB: Part II}

\hypersetup{
pdftitle={A template for the arxiv style},
pdfsubject={q-bio.NC, q-bio.QM},
pdfauthor={David S.~Hippocampus, Elias D.~Striatum},
pdfkeywords={First keyword, Second keyword, More},
}

\begin{document}
\maketitle

\begin{abstract}
Full-waveform inversion (FWI) requires accurate initial velocity models to avoid cycle-skipping, but constructing such models remains challenging in practice. Building on the depth-progressive diffusion framework introduced in Part~I, which relied on idealized reflectivity constraints, this work adapts the methodology to realistic exploration scenarios. We replace perfect structural information with migration-derived attributes extracted from seismic images, and introduce smooth background velocity models from tomography as additional conditioning inputs. The framework jointly leverages background/migration velocity, migrated structural information, and sparse well measurements to synthesize high-resolution velocity models through depth-progressive generation. Validation on synthetic examples demonstrates superior accuracy compared to conventional interpolation and alternative deep learning methods, with generated models successfully initializing FWI and mitigating cycle-skipping even in complex geological structures. Field data confirms practical applicability: despite training on synthetic data, the method generalizes effectively to field conditions, producing velocity models with synthetic data response that nearly match observed seismic data. As a result, this framework establishes a practical pathway to deploy generative diffusion models for velocity model building under realistic constraints.
\end{abstract}

\section{\textbf{Introduction}}

Full-waveform inversion (FWI) reconstructs subsurface velocity by minimizing the mismatch between observed and simulated wavefields, enabling recovery of fine-scale structures that are beyond the reach of traveltime methods. With multiscale frequency strategies and appropriate regularization, FWI can retrieve high-wavenumber details while respecting the large-scale kinematics features of the velocity model, leading to improved imaging and more reliable subsurface interpretation \citep{tarantola1984inversion, bunks1995multiscale, virieux2009overview}. Despite its potential, FWI remains susceptible to nonconvexity and cycle skipping, especially when low frequencies are limited or the starting model lacks the correct long-wavelength trend \citep{operto2013guided}. In such cases, phase misalignment at early iterations can drive the optimization toward local minima, corrupting deep updates and degrading the final result. Consequently, the quality of the initial or background velocity model is a critical determinant of FWI success.

Addressing this challenge requires methods that can 
build plausible velocity models while incorporating 
available constraints and respecting subsurface structure. To construct such high-quality velocity models, in the companion paper (Part~I), we introduced a depth-progressive diffusion framework that synthesizes velocity from shallow to deep by propagating reliable near-surface information. There, structural constraints were provided by a vertical reflectivity model that was assumed to be known. This reflectivity acted as a perfect proxy for subsurface layering and faults, and was directly used to guide the diffusion model during depth-progressive synthesis. While this setup is well suited for testing the conceptual behavior of the method, it is not representative of practical exploration realities, where the true reflectivity is unavailable.

In practice, the structural information available for velocity model building (VMB) should be inferred indirectly from seismic data. A natural starting point is the conventional VMB workflow, in which a smooth background/migration velocity model is constructed through velocity analysis and tomography \citep{alkhalifah2003tau, sava2004wave}. If well logs are available, the background can be further refined by well-guided, dip-consistent interpolation \citep{chen2016geological, chen2019interpolation, huang2020geological, li2022well, zhang2023structurally}. In these approaches, the background model is first used to migrate the data and estimate structural dips from the migrated image. The well-log velocities are then interpolated along reflector directions to fill the model between wells. This workflow preserves the low-wavenumber trend, respects the sparse but reliable well constraints, and introduces layer-parallel variations where the data are well illuminated. However, its performance strongly depends on having several wells that are reasonably distributed. In reality, drilling is expensive, so wells are few and often clustered. As a result, velocity estimates far from the wells depend mainly on the slope field and smoothing assumptions, which may spread biases across faults, perform poorly in areas with weak illumination, and leave the deep velocity trend poorly constrained with increasing uncertainty.

Despite these limitations, the workflow suggests useful insight: we may relax the requirement for an accurate structural model. Instead of relying on a true reflectivity model, we can first migrate the data with a conventional background/migration velocity (from velocity analysis and tomography) to obtain a relatively low resolution but structurally informative image to represent the reflectivity. The structural dips and reflector geometry approximately embedded in this image can then serve as a practical surrogate for the ideal reflectivity used in Part~I. In this way, the diffusion framework is conditioned on image-derived structure rather than on an unavailable ground-truth reflectivity, allowing us to operate under realistic acquisition, illumination, and well-control conditions. Even if the quality of the image is poor at depth, which is often the case in realistic scenarios, the depth progression nature of the framework, as we saw in Part I, will help regularize the estimation. 

Building on this idea, in this Part~II, we adopt a more realistic constraint design. We first obtain a migrated image from a smooth background/migration model. We then jointly condition the depth-progressive diffusion on (i) sparse well constraints, (ii) migration-derived structural information, and (iii) the migration velocity. The jointly conditioned synthesis produces velocity models intended as high-quality FWI initializations, with the goal of reducing phase mismatch, mitigating cycle skipping, and improving the recovery of deep features under realistic acquisition and noise conditions. We evaluate the workflow using synthetic examples and a marine field data set to assess its effectiveness under realistic considerations.
\section{\textbf{Brief recap of the depth-progressive diffusion framework}}
\begin{figure}[htbp]
\centering
\includegraphics[width=1\textwidth]{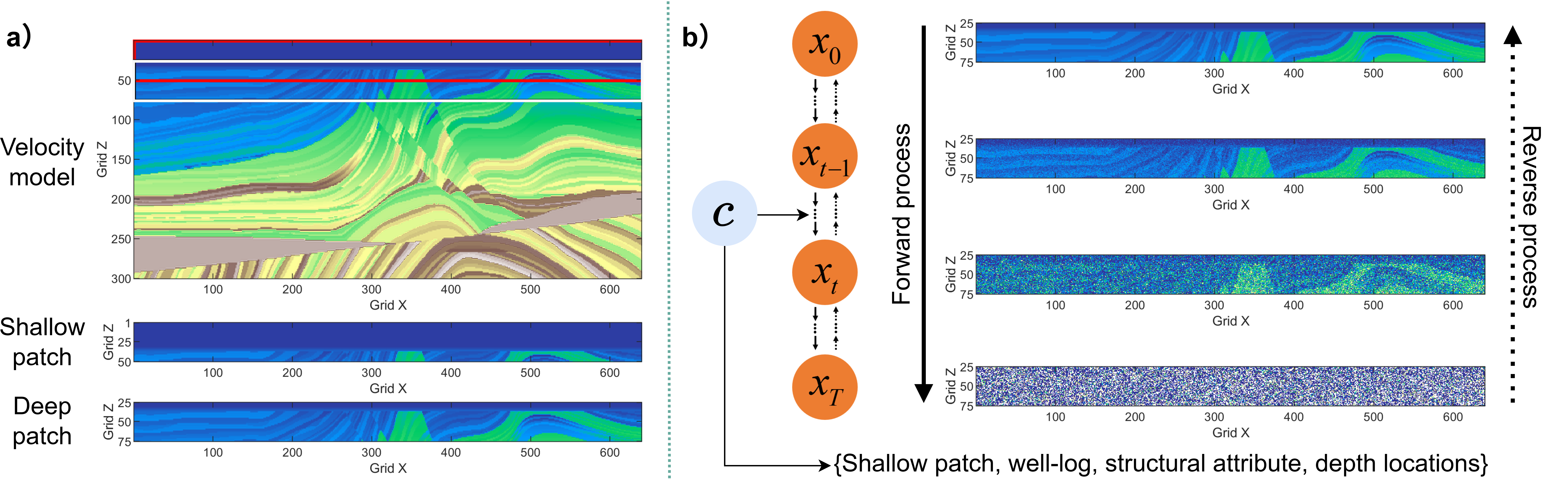}
\caption{Illustration of the depth-progressive diffusion framework from Part~I. (a) Training data extraction: overlapping shallow (red box) and deep (white box) velocity windows are sampled from full-scale models. (b) Diffusion process: the forward process progressively adds noise to the clean deep patch $x_0$, producing noisy states up to $x_T$. The reverse process learns to denoise $x_t$ back to $x_0$ conditioned on the preceding depth shallow patch, well-log, structural attributes (true reflectivity in Part~I), and the patches depth locations. }
\label{fig1}
\end{figure}

Part~I introduced a depth-progressive diffusion framework that constructs seismic velocity models from shallow to deep by explicitly propagating shallow priors.  Figure~\ref{fig1} illustrates the key components of this framework. Instead of generating an entire velocity model, the method synthesizes overlapping depth windows sequentially so that each deeper window is conditioned on the previously generated shallower window. This design aligns model construction with the top-to-down information flow of surface seismic data.

During training, pairs of shallow-deep windows are sampled from full-scale models, as illustrated in Figure~\ref{fig1}a. The two windows overlap in depth, with the overlap size randomly varied during training to improve robustness. The deeper window (indicated by the white box) serves as the denoising target $x_0$, while the shallower window (indicated by the red box) provides a velocity prior. The forward diffusion process (Figure~\ref{fig1}b) progressively corrupts $x_0$ with Gaussian noise to produce increasingly noisy states $x_1, x_2, \ldots, x_T$. The network learns the reverse process: given a noisy sample $x_t$ and conditioning information $\mathbf{c}$, it predicts the clean velocity patch. The conditioning includes the shallow velocity window, well-log velocities, structural attributes (reflectivity in Part~I), and absolute depth encodings for both windows. These depth encodings explicitly inform the network of each grid point's depth, enabling it to learn depth-dependent velocity characteristics. To handle scenarios with incomplete information, a mixed conditional-unconditional training strategy randomly drops well and structural constraints with small probability, teaching the model to generate plausible velocities even when certain inputs are unavailable.

At inference, the model begins from a shallow prior (like the water layer for marine data) and proceeds downward with a fixed stride so that consecutive windows overlap. Each deeper window is predicted while being conditioned on the previously generated window and the depth encodings, with optional well and structural inputs when available. Overlaps are merged by Gaussian-weighted accumulation rather than simple concatenation, where the running weights are normalized at the end. This fusion suppresses grid artifacts, favors the better-constrained window centers, and yields a coherent velocity section. Because the diffusion process is probabilistic, multiple realizations can be generated by sampling different Gaussian noise seeds for each window while holding the conditioning fixed. Ensemble statistics computed per window and after fusion provide mean and variance fields that highlight underdetermined regions and offer a practical measure of VMB uncertainty.

Part~I demonstrated that shallow-to-deep conditioning, absolute depth encodings, and Gaussian fusion across overlaps are key to stable synthesis, and that mixed conditioning improves tolerance to sparse or missing constraints. Building on these elements, Part~II adapts the framework to realistic settings by replacing ideal reflectivity with a reverse time migration (RTM) image and by incorporating the migration velocity available from conventional VMB workflow, to produce migration velocities used in the imaging.

\section{\textbf{Depth-progressive velocity model synthesis under realistic considerations}}

In this section, we adapt the framework to operate under realistic conditions by replacing the ideal reflectivity constraint with an RTM image representation of the reflectivity as well as incorporating the often smooth migration velocity model as an additional conditioning input.

\subsection{Migration-based constraints}

The key distinction between Part~I and the present work lies in how we obtain and utilize structural information. Rather than assuming access to true reflectivity, we adopt a workflow that mimics conventional velocity model building practices:

\begin{enumerate}
    \item \textbf{Migration velocity construction}: We begin with a smooth background velocity model $v_{mig}$ obtained through conventional velocity analysis and tomography \citep{alkhalifah2003tau, sava2004wave}. This model captures the large-scale, low-wavenumber velocity trend but lacks fine-scale structural details. In our synthetic experiments, we simulate this background by applying strong gaussian smoothing to the true velocity model. For field data applications, the background comes directly from standard migration velocity analysis (MVA) workflows.
    
    \item \textbf{Migration-based structural information}: The background model $v_{mig}$ is used to migrate the seismic data, producing an image $s_{mig}$. While this migrated image is imperfect, where it may contain positioning errors, illumination gaps, and migration artifacts, it nevertheless provides valuable information about subsurface structure. Within this image, structural attributes are embedded such as local dip fields, reflector continuity, and edge-detected boundaries. These attributes, extracted from the network, serve as practical surrogates for the true reflectivity used in Part~I.
    
    \item \textbf{Joint conditioning strategy}: During both training and inference, the diffusion model is now conditioned on three complementary sources of information:
    \begin{itemize}
        \item \textbf{Shallow velocity priors} ($v_{shallow}$): Propagated from previously generated depth windows, providing vertical continuity.
        \item \textbf{Structural information} ($s_{mig}$): Extracted from the migrated image, encoding reflector geometry and geological layering.
        \item \textbf{Background velocity trend} ($v_{mig}$): Extracted from the migration velocity, providing the low-wavenumber reference that ensures generally consistent velocity synthesis.
        \item \textbf{Localized ground truth} ($w$): Provided by the sparse but reliable well measurements where available.
    \end{itemize}
\end{enumerate}

This joint conditioning strategy allows the diffusion model to respect the large-scale velocity trend from tomography, honor the sparse well data, and introduce geologically plausible fine-scale variations guided by the migration-derived structure. The conditioning tensor at each depth window $i$ is thus:
\begin{equation}
    \mathbf{c}_i = \{v_{shallow}, d_{shallow}, d_{deep}, v_{mig,i}, s_{mig,i}, w_i\},
\end{equation}
where $v_{mig,i}$ and $s_{mig,i}$ denote the migration velocity and image, respectively corresponding to depth window $i$, and $d_{shallow}$ and $d_{deep}$ represent the depth positions of grid points for shallow and deep patches, respectively.

Importantly, unlike Part~I where the structural constraint was perfect, the migration image $s_{mig}$ is often noisy, possibly inaccurate positioning and poorly illuminated. The diffusion model is expected learn to leverage this imperfect structural guidance while avoiding overfitting to migration artifacts. This is achieved through the training strategy described in the next subsection, where the model is exposed to realistic background-migration pairs during training.

\subsection{Training dataset construction under realistic acquisition scenarios}

To train the diffusion model for realistic VMB, we must construct a training dataset that reflects the characteristics of practical exploration workflows. This involves generating triplets of (i) true velocity models $v_{true}$, (ii) smooth background/migration velocity models $v_{mig}$, and (iii) migration images $s_{mig}$. The construction process is designed to mimic practical exploration workflows and account for realistic acquisition conditions.

The target velocity models $v_{true}$ serve as the ground truth for training and should reflect geologically plausible subsurface structures relevant to the exploration area of interest. When well logs are available in the survey area, they provide valuable guidance for velocity model construction by informing us of the approximate velocity ranges and depth trends in the subsurface. These well-derived information, as well as our geological prior information, serve to guide the construction of geologically reasonable velocity models. In the absence of well data, we rely on our experience and expectations of the region, which includes approximate velocity ranges, depth trends, and structural styles typical of the geological setting. Existing model databases, regional velocity studies, or conceptual geological models provide guidance for generating plausible velocity distributions. The goal is to create a diverse set of training models that span the range of geological complexity and velocity variations expected in the target exploration area.

For each target velocity model $v_{true}$, we compute a corresponding smooth background model $v_{mig}$ by applying a spatial smoothing operator $v_{mig} = \mathcal{S}(v_{true})$, where $\mathcal{S}$ denotes a smoothing operation that removes fine-scale features while preserving long-wavelength trends. The smoothing is not arbitrary, as it should reflect the resolution and smoothness characteristics of velocity models obtained from real tomographic workflows. However, determining the exact smoothing parameters that match real tomographic resolution is often challenging. Therefore, to account for this discrepancy, we can relax the smoothing requirement by allowing the smoothing parameters (e.g., Gaussian kernel widths) to vary within a reasonable range during training, ensuring that the diffusion model learns to handle different levels of background quality. An alternative guide is provided by a smoothing measure, like the total variation measure of the available migration velocity, which can serve as guidance to the amount of smoothing we can use in preparing the training set.

To obtain realistic structural constraints $s_{mig}$, we simulate the seismic acquisition and imaging workflow used in the field data. Using the true velocity model $v_{true}$, we perform forward modeling to generate synthetic shot gathers, where the acquisition geometry (source and receiver positions, sampling intervals) is designed to closely match the configuration of the actual field survey. To further align the synthetic data characteristics with field data, we can extract the source wavelet from the actual seismic data using some wavelet estimation methods, ensuring that the frequency content, bandwidth, and phase characteristics of the synthetic data closely resemble those of the field observations. The synthetic shot gathers are then migrated using the smooth background velocity model $v_{mig}$ to produce the migrated image $s_{mig}$. Because $v_{mig}$ differs from the true velocity $v_{true}$, the resulting image may contain positioning errors and artifacts similar to those encountered in real migration workflows. 

These strategies ensure that the training dataset captures the characteristics of real seismic imaging, and matching the acquisition geometry and source wavelet to the target field survey ensures that the energy distribution and structural features in the training images align with those in actual field data, enabling better generalization during inference.

\subsection{Training strategy and network adaptation}

To effectively incorporate the migration velocity model as an additional conditioning constraint, we extend the training strategy and network architecture from Part~I. The training continues to employ a mixed conditional and unconditional approach to handle scenarios with incomplete conditioning information during inference. Following the same strategy used for structural and well constraints in Part~I, the migration velocity model $v_{mig}$ is randomly dropped with a probability of 5\% during training. 

Meanwhile, the introduction of the migration velocity constraint necessitates corresponding modifications to the network architecture. Similar to how structural constraints are encoded in Part~I, we introduce a dedicated migration velocity embedding layer, which follows a similar architecture as the structure embedding layer, to effectively integrate $v_{mig}$ into the generation process. Also, to incorporate the encoded migration velocity information into the generation process, we add an additional branch in each residual block of the U-Net backbone, analogous to the treatment of structural constraints in Part~I. Each residual block now receives three types of encoded conditioning information: well velocities, migration images, and migration velocity models. As a result, we can allow the network to jointly leverage all available conditioning information throughout the denoising process, enabling effective integration of the background velocity trend with fine-scale structural guidance and sparse well constraints.
\section{\textbf{Synthetic examples}}\label{sec:synthetic_example}
In the following, we first validate our depth-progressive velocity synthesis framework under realistic considerations using two synthetic examples. 

\subsection{Training dataset}\label{subsec:synthetic_dataset}

We adopt the training dataset from \cite{zhang2025well}, which is specifically designed to simulate realistic exploration scenarios with limited well control. The dataset is derived from a modified Marmousi velocity model with dimensions of $300\times640$ grid points and a uniform grid spacing of 12.5~m. Three vertical velocity profiles are extracted at horizontal positions of 1.25, 3.75, and 6.5~km to serve as hypothetical well logs. For each well location, 100 synthetic velocity realizations are generated through a well-guided construction process, producing a total of 300 training models. To increase geological complexity and better represent realistic subsurface conditions, artificial faults are introduced into these synthetic models, ensuring that each realization exhibits localized well constraints while maintaining geologically plausible lateral and vertical velocity variations.

The acquisition geometry is designed to reflect challenging field conditions with limited offset coverage. For each velocity model, 106 shots are simulated using the acoustic wave equation, with sources uniformly distributed along the surface. Each shot is recorded on a single-sided array of 120 receivers with offsets ranging from 0.25 to 3.0~km, representing conventional towed-streamer marine surveys. A Ricker wavelet with a dominant frequency of 20~Hz serves as the source signature. The smooth background velocity models $v_{mig}$ are generated by applying strong Gaussian smoothing ($\sigma=25$ grid points) to each true velocity model, mimicking the resolution limitations of conventional tomographic methods. These background/migration velocity models are then used to perform acoustic reverse-time migration (RTM) on the synthetic shot gathers, producing the migration images $s_{mig}$ that serve as conditioning inputs during training and inference.

\begin{figure}[htbp]
\centering
\includegraphics[width=1\textwidth]{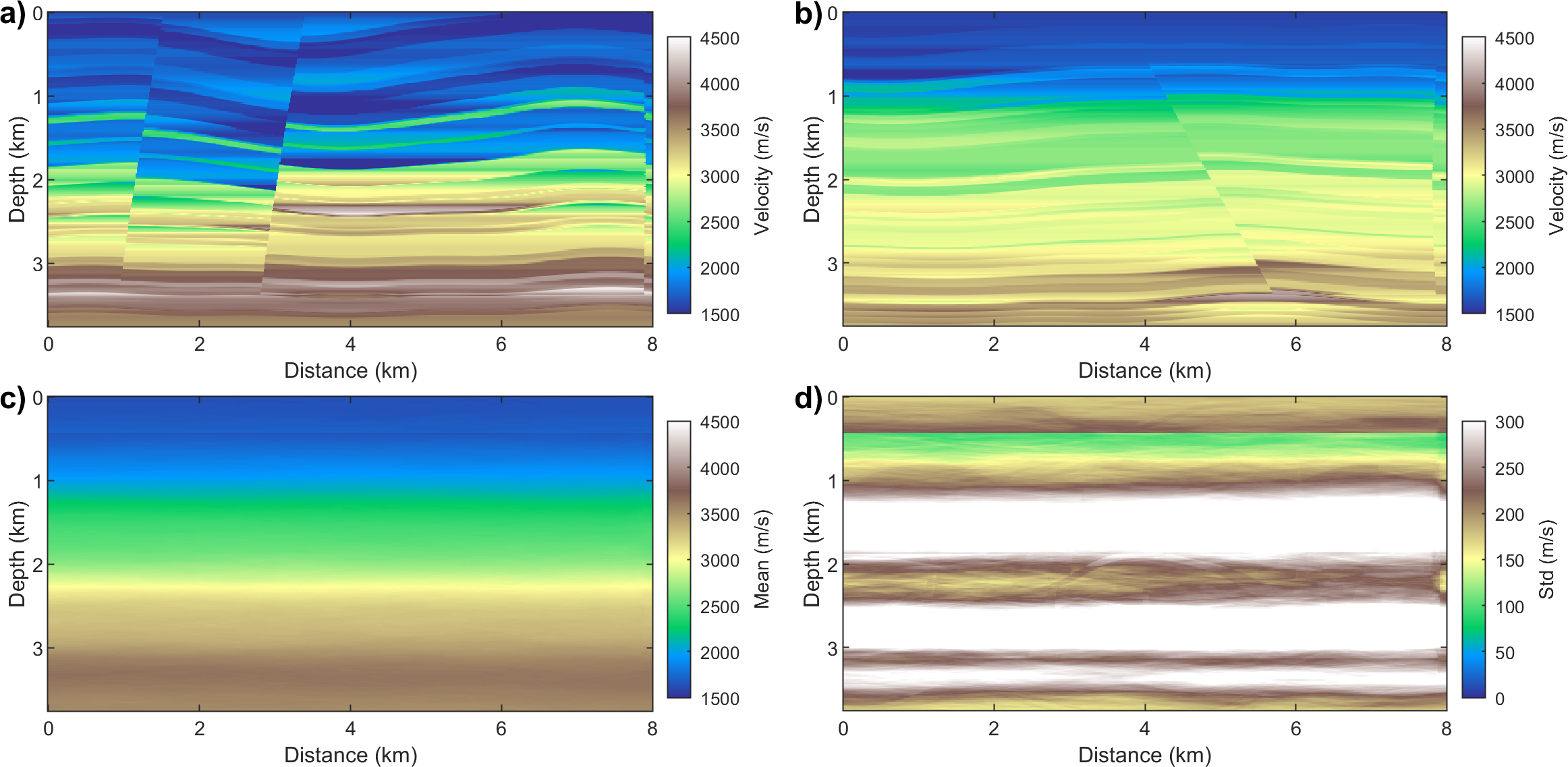}
\caption{Statistical characteristics of the augmented training dataset consisting of 600 synthetic velocity models. (a) and (b) are two representative sample velocity models. (c) Mean velocity model. (d) Standard deviation map.}
\label{fig2}
\end{figure}

To simulate realistic well scenarios, only a single well is randomly selected from the true velocity models during each training iteration. This strategy ensures that the trained diffusion model learns to synthesize velocity models under conditions of limited well coverage, which is representative of practical exploration scenarios where drilling is expensive and well spacing is typically large. If the well location or locations for the target real dataset are known, we can use the same well locations in the training. To augment the training dataset, we apply horizontal flipping to all velocity models and their corresponding migration images, effectively doubling the dataset size to 600 training samples. Figure~\ref{fig2} illustrates representative samples and the statistical characteristics of the augmented training dataset. The mean velocity model (Figure~\ref{fig2}c) reveals the general subsurface structure captured by the ensemble, showing a progressive velocity increase with depth. The standard deviation map (Figure~\ref{fig2}d) quantifies the diversity within the training set, with higher variability concentrated in regions, like 1-2 km. These statistical characteristics reflect the prior knowledge embedded in the training dataset, which will influence the velocity models synthesized by the diffusion model, as illustrated in the subsequent results.

Figure~\ref{fig2} illustrates representative examples and statistical characteristics of the augmented training dataset. Figures~\ref{fig2}a,b show two synthetic velocity models from the training set, exhibiting the range of geological complexity and structural variability present in the dataset. The ensemble mean (Figure~\ref{fig2}c) reveals the general subsurface structure captured by the training set, showing the overall velocity trend with depth. The standard deviation map (Figure~\ref{fig2}d) quantifies the diversity within the training set, with higher variability concentrated in regions such as 1-2~km depth where geological uncertainty is largest. These statistical characteristics reflect the prior knowledge embedded in the training dataset, which will influence the velocity models synthesized by the diffusion model, as illustrated in the subsequent results.

\subsection{Benchmark methods and training configuration}

To evaluate the effectiveness of our depth-progressive framework, we compare it against two benchmark approaches: a deep learning (DL)-based method and a conventional VMB workflow.

The first benchmark (denoted by GenAIB) is the diffusion-based VMB method proposed by \cite{zhang2025well}, which serves as a representative DL benchmark for FWI initialization. \cite{zhang2025well}'s approach, which is based on controllable diffusion velocity model synthesis of \cite{wang2024controllable}, also employs GDMs to construct initial velocity models for FWI, but differs fundamentally in its formulation. Their method treats velocity model generation as a conditional image generation problem, where the network directly generates the entire velocity model in a single pass. During training, the conditioning information, including the initial velocity model, the migration image derived from the initial model, and well constraints, is concatenated with the noised version of the true velocity model and fed into the network. The network is trained to predict the clean velocity model through the reverse diffusion process. A key limitation of their approach is that it does not incorporate unconditional training, meaning the network cannot handle scenarios where conditioning information is unavailable. Despite this difference in conditioning strategy, their method provides an appropriate benchmark as it shares the same objective of using diffusion models for VMB.

The second benchmark (denoted by WDIB) is the well-guided, dip-consistent interpolation method \citep{chen2016geological, li2022well}, which represents the conventional approach introduced in our Introduction section. This workflow first migrates the seismic data using the smooth background velocity model to produce a image, from which structural dip fields are extracted. The well-log velocities are then interpolated along the estimated reflector directions to construct a refined velocity model that honors both the sparse well constraints and the structural geometry. This method has been widely adopted in industry and provides a practical baseline for assessing whether data-driven approaches offer improvements over established workflows.

This DL-based WDIB method and our method are trained on the same dataset described above. The network architectures are based on U-Net backbones with matching depth and feature dimensions at each resolution level, ensuring a fair comparison in terms of model capacity. The primary architectural differences lie in the residual block design and the conditioning embedding mechanism, which in our case accommodates the mixed conditional-unconditional training strategy. For our depth-progressive method, we use a batch size of 24, the AdamW optimizer with a fixed learning rate of 1e-4, and exponential moving average (EMA) with a rate of 0.999 to stabilize training, consistent with the configuration in Part~I. The model is trained for 100000 iterations, requiring approximately 27 hours on a single NVIDIA A100 GPU. For \cite{zhang2025well}'s benchmark method, we replicate their training procedure for 100000 iterations under the same optimizer and learning rate settings. However, due to the higher memory consumption of their architecture (which processes the full model in a single pass), the batch size is reduced to 8. The training completes in approximately 31 hours on the same hardware.

\subsection{Synthetic velocity model within the distribution}

\begin{figure}[htbp]
\centering
\includegraphics[width=1\textwidth]{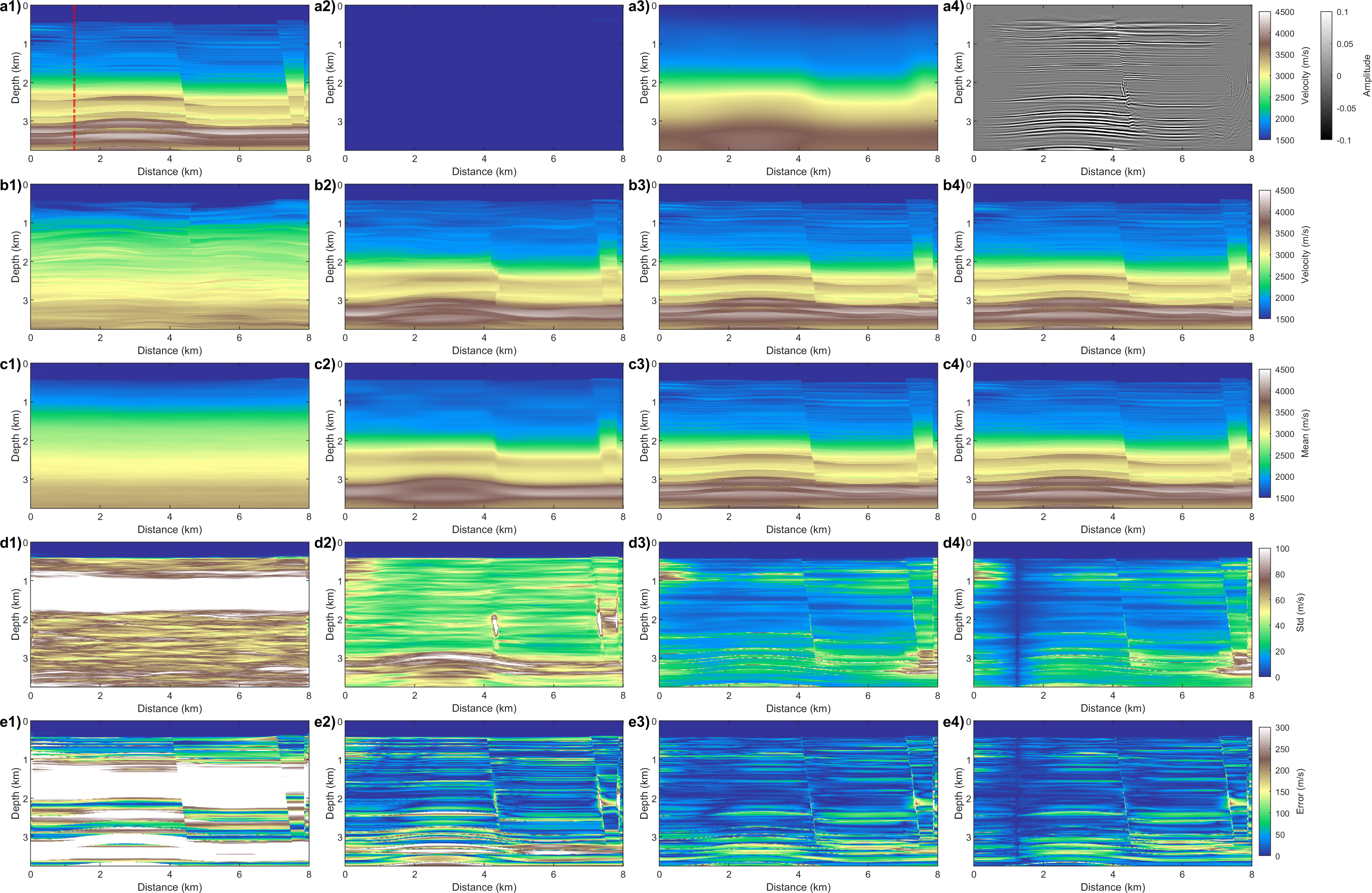}
\caption{Velocity model building results for an in-distribution synthetic model under different conditioning scenarios. First row (a1-a4): true velocity model with well location (red dashed line), shallow prior, background velocity model, and migration image. Second row (b1-b4): individual realizations generated under unconditional, background-only, background+image, and background+image+well conditioning. Third row (c1-c4): the mean of the 50 realizations for the four scenarios. Fourth row (d1-d4): the standard deviation of the 50 realizations. Fifth row (e1-e4): absolute error between the mean (row c) and true model (a1).}
\label{fig3}
\end{figure}

We begin our test on a synthetic velocity model drawn from the training distribution. Figure~\ref{fig3} presents the velocity VMB results under unconditional and various conditional settings. The first row (a1-a4) displays, from left to right, the true velocity model, the shallow prior used to initialize the depth-progressive generation, the smooth background/migration velocity model used for migration, and the corresponding migrated image, respectively. The second row (b1-b4) shows individual realizations generated under four conditioning scenarios: unconditional, background velocity only, both background velocity and migration image, and all three constraints (background velocity, image, and well), respectively. The third row (c1-c4) presents the mean of 50 generated realizations for each scenario. The fourth row (d1-d4) displays the corresponding standard deviation maps, which quantify VMB uncertainty. Finally, the fifth row (e1-e4) shows the absolute error between the mean and the true velocity model.

\begin{figure}[htbp]
\centering
\includegraphics[width=1\textwidth]{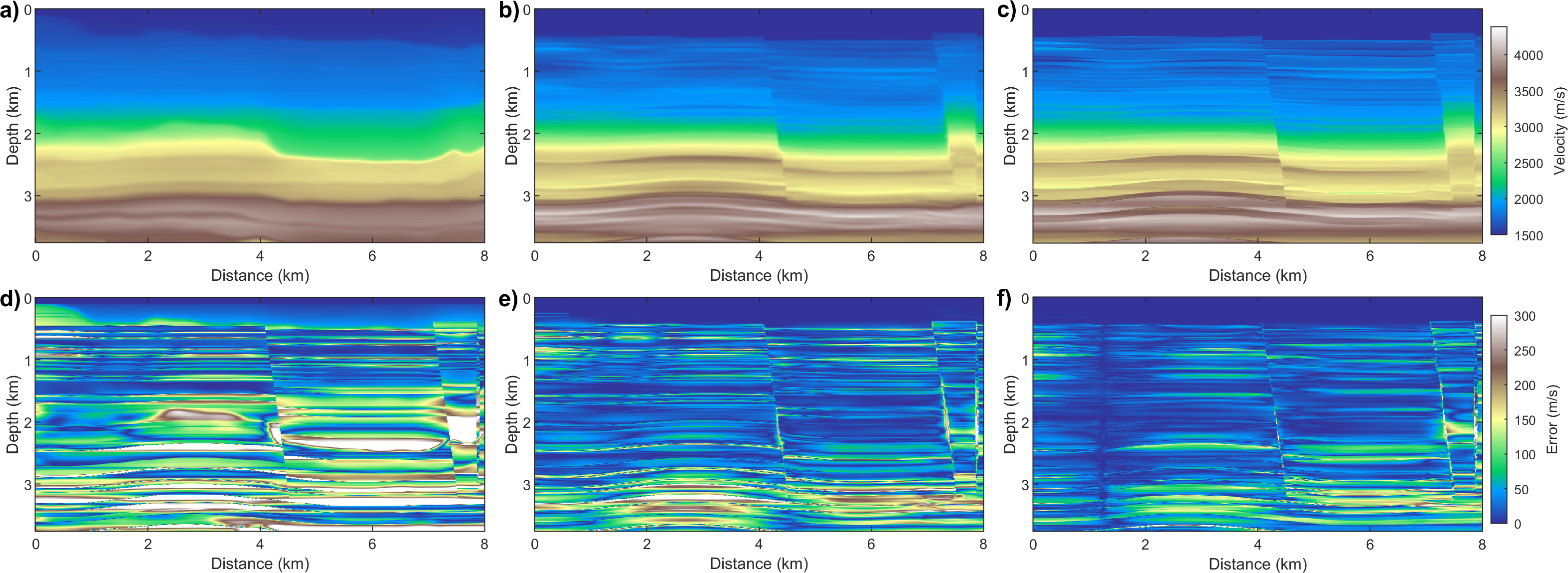}
\caption{Comparison of velocity model building results for the in-distribution synthetic model. First row (a-c): velocity models generated by WDIB, GenAIB, and our method (the mean of the 50 realizations). Second row (d-f): corresponding absolute errors relative to the true velocity model.}
\label{fig4}
\end{figure}

Under the unconditional setting (first column), our framework successfully generates high-resolution velocity models, influenced by the training set, exhibiting a progressive velocity increase with depth, starting from the shallow prior. However, without additional constraints, the individual realization (Figure~\ref{fig3}b1) does not align well with the true velocity model. The mean (Figure~\ref{fig3}c1) closely resembles the training dataset mean (Figure~\ref{fig2}a), confirming that unconditional generation naturally falls within the learned prior distribution. Similarly, the standard deviation map (Figure~\ref{fig3}d1) exhibits a spatial pattern consistent with the training dataset variability (Figure~\ref{fig2}b), with notably high uncertainty in the 1-2~km depth range. The error map (Figure~\ref{fig3}e1) reveals substantial errors, particularly in regions where uncertainty is highest. This correspondence between uncertainty and error demonstrates that the uncertainty quantification provided by our probabilistic framework serves as a reliable indicator of VMB accuracy.

Introducing the background velocity constraint (second column) yields a significant improvement in VMB accuracy. The generated models now broadly match the true velocity structure, as evidenced by both the individual realization (Figure~\ref{fig3}b2) and the mean (Figure~\ref{fig3}c2). However, due to the smooth nature of the background velocity, the generated models exhibit limited resolution, particularly noticeable in the mean (Figure~\ref{fig3}c2) which appears overly smooth. The standard deviation map (Figure~\ref{fig3}d2) effectively identifies regions of high uncertainty, such as structurally complex fault zones and deep high-velocity areas. The error map (Figure~\ref{fig3}e2) confirms a substantial reduction in VMB error compared to the unconditional case. Importantly, regions with larger errors correlate well with regions with higher standard deviation, further validating the reliability of the uncertainty estimates.

Further incorporating migration images (third column) leads to additional improvements in both accuracy and resolution. The generated velocity model (Figure~\ref{fig3}b3) now exhibits significantly enhanced resolution, with structural features that align well with the true model. The mean (Figure~\ref{fig3}c3) clearly captures fine-scale layering and fault geometries, indicating that migration-derived structure provides effective guidance for introducing medium-to-high wavenumber information while respecting geological structure. An additional benefit is the reduction in VMB uncertainty (Figure~\ref{fig3}d3), suggesting that structural constraints narrow the range of plausible velocity variations. Nevertheless, structurally complex fault zones and deep high-velocity regions remain the most challenging areas (Figure~\ref{fig3}e3).

Adding well constraints on top of the background and structural conditions (fourth column), provided by the migration velocity and the migration image, respectively, yield improved local control near well locations. While the individual realization (Figure~\ref{fig3}b4) and the mean (Figure~\ref{fig3}c4) do not exhibit obvious visual differences compared to the case without wells, the standard deviation (Figure~\ref{fig3}d4) and error maps (Figure~\ref{fig3}e4) reveal clear benefits. The uncertainty and error are notably reduced near the well (marked by the red dashed line in Figure~\ref{fig3}a1), demonstrating the unique role of sparse but accurate well measurements in enhancing the velocity estimates locally. This validates the effectiveness of our joint conditioning strategy in integrating complementary information sources, i.e., background velocity for large-scale trends, migration structure for intermediate-scale layering, and wells for local accuracy.

\begin{figure}[htbp]
\centering
\includegraphics[width=1\textwidth]{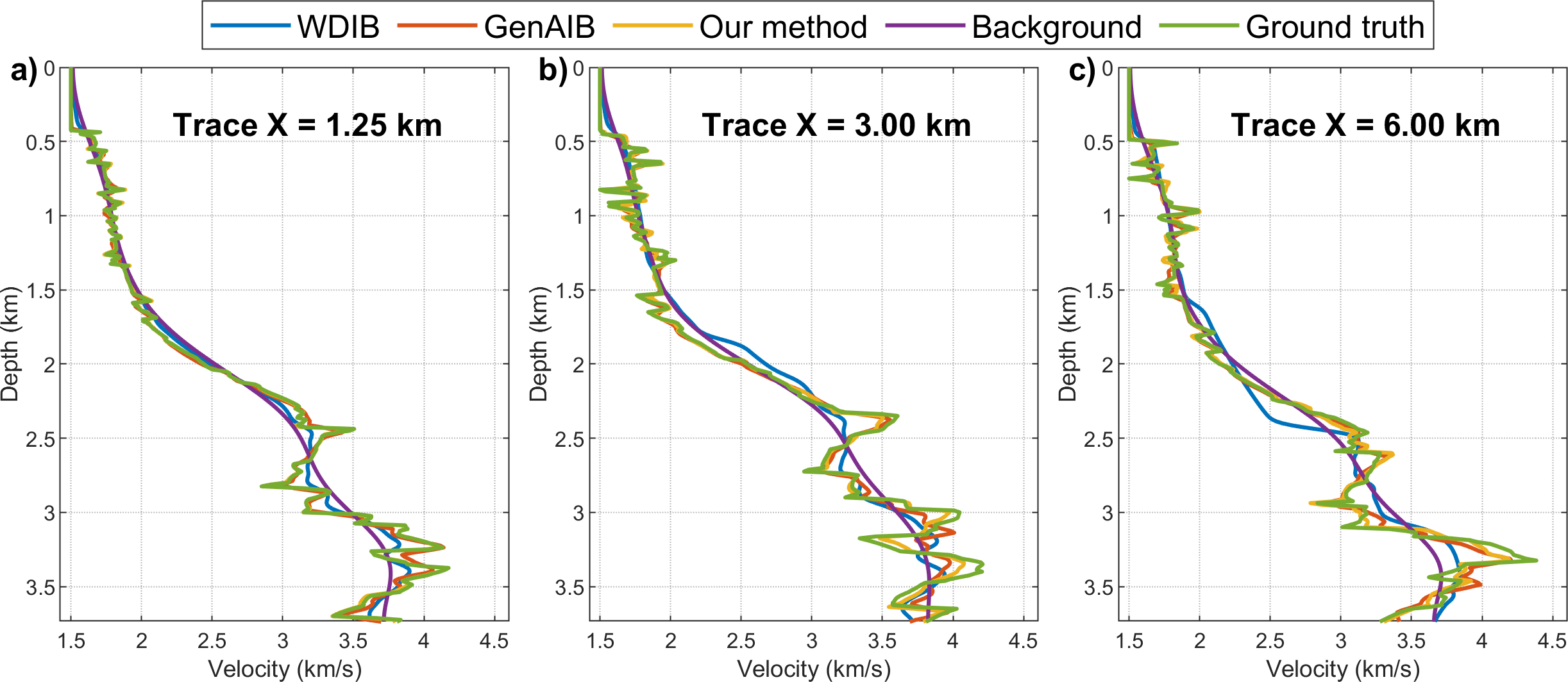}
\caption{Vertical velocity profiles for the in-distribution synthetic model extracted at three lateral positions: (a) 1.25~km (well location), (b) 3.00~km, and (c) 6.00~km. The profiles compare WDIB (blue), GenAIB (orange), our method (yellow), background velocity (purple), and ground truth (green).}
\label{fig5}
\end{figure}

We next compare our framework against the two benchmark methods, WDIB and GenAIB. Both DL-based methods use all available constraints, with results showing the mean of 50 realizations. Figure~\ref{fig4} presents the VMB results and corresponding errors. The first row (a-c) shows the velocity models generated by WDIB, GenAIB, and our method, while the second row (d-f) displays the absolute errors. We can see that both DL-based methods significantly outperform the conventional WDIB approach (Figure~\ref{fig4}a,d), which exhibits limited resolution and notable errors, particularly in regions distant from wells and in structurally complex areas. This reflects the inherent limitation of interpolation methods, which rely heavily on smoothness assumptions and struggle to capture fine-scale velocity variations. In contrast, GenAIB (Figure~\ref{fig4}b) produces substantially improved results with better resolution and structural fidelity. Our method (Figure~\ref{fig4}c) achieves further improvement, as evidenced by the reduced error in Figure~\ref{fig4}f. A particularly notable difference is the effectiveness with which our method honors the well constraint, where the VMB error near the well location (around 1.25~km) is noticeably lower for our method, indicating that our framework more effectively integrates local well measurements with regional structural and background velocity constraints. 

\begin{table}[h]
\centering
\caption{Quantitative comparison of velocity model building accuracy for the in-distribution synthetic model. MAE (mean absolute error) is reported in m/s, and SSIM (structural similarity index) ranges from 0 to 1, with bold text indicating best accuracy.}\label{tab1}
\begin{tabular}{lccc}
\toprule
 & WDIB & GenAIB & Our method \\
\midrule
MAE & 100.67 & 49.27 & \textbf{33.91} \\
SSIM & 0.767 & 0.872 & \textbf{0.942} \\
\bottomrule
\end{tabular}
\end{table}

To provide a more detailed comparison, Figure~\ref{fig5} presents vertical velocity profiles extracted at three representative lateral positions: 1.25~km (well location), 3.0~km, and 6.0~km. These profiles compare the results from WDIB (blue), GenAIB (orange), our method (yellow), background velocity (purple), and ground truth (green). At the well location (Figure~\ref{fig5}a), all methods align reasonably well in shallow depths, but our method and GenAIB maintain closer agreement at greater depths. At 3.0~km (Figure~\ref{fig5}b), both DL-based methods track the true velocity more accurately than WDIB, with our method showing slightly better fidelity in the deeper section below 3.0~km. At 6.0~km (Figure~\ref{fig5}c), where well control is distant, our method continues to demonstrate superior performance. Table~\ref{tab1} provides quantitative assessment by calculating the mean absolute error (MAE) and structural similarity index (SSIM) between the generated velocity models and the true model. Our approach achieves the lowest MAE (33.91~m/s vs. 49.27 for GenAIB and 100.67 for WDIB) and highest SSIM (0.942 vs. 0.872 and 0.767), confirming superior accuracy.

\begin{figure}[htbp]
\centering
\includegraphics[width=1\textwidth]{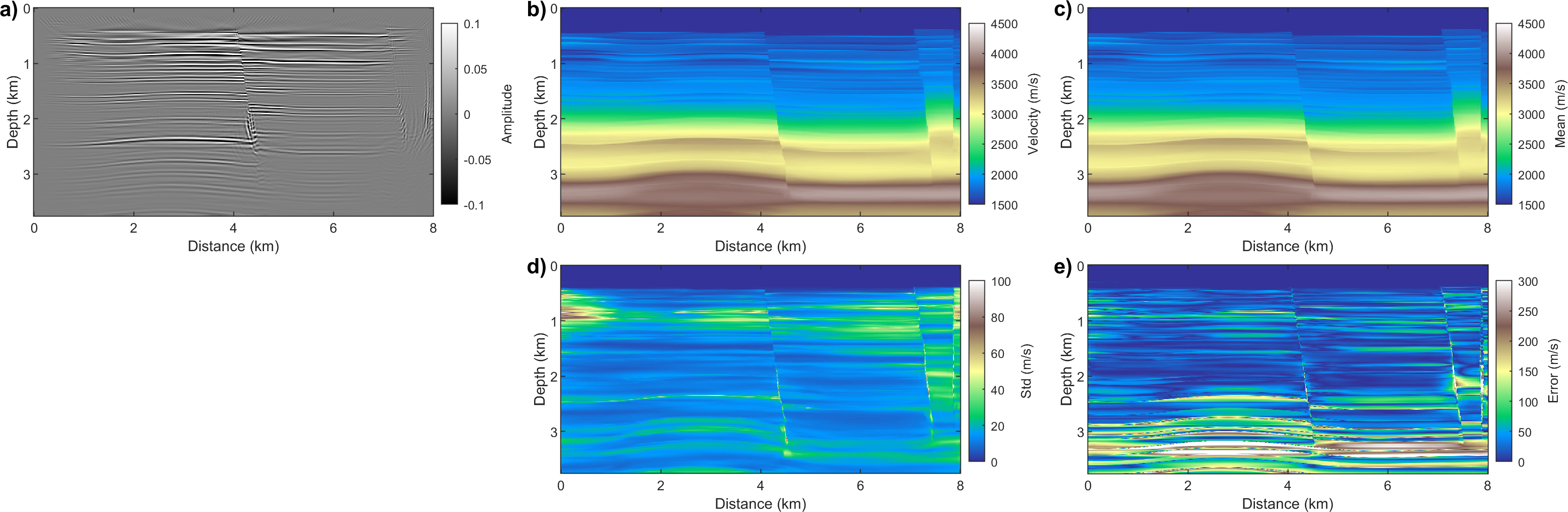}
\caption{Velocity model building results with attenuated deep part of the image providing limited structural constraints. (a) Migration image with energy artificially attenuated below 2.5~km to simulate poor deep illumination (compare with Figure~\ref{fig3}a4). (b) A realization generated under background+attenuated structure conditioning. (c) The mean of the 50 realizations. (d) The standard of the 50 realizations. (e) Absolute error between the mean (c) and true model (Figure~\ref{fig3}a1).}
\label{fig6}
\end{figure}

To highlight the power and role of the shallow to deep progression, we repeat the test in Figure \ref{fig3}, but with the deep part of the migration image are artificially attenuated to emulate real conditions. Figure~\ref{fig6} presents results when the migration image is attenuated beyond 2.5~km depth to replicate scenarios where deep areas are poorly illuminated due for example to high attenuation. Figure~\ref{fig6}a shows the attenuated migration image, where structural information below 2.5~km is severely degraded compared to the original full-quality image in Figure~\ref{fig3}a4. We condition the generation on both background/migration velocity and this weakened structural constraint. Compared with the result using the original strongly illuminated migration image (Figure~\ref{fig3}b3,c3), we can observe that both the individual realization (Figure~\ref{fig6}b) and mean (Figure~\ref{fig6}c) exhibit reduced resolution in the deep area below 2.5~km, and the VMB accuracy shows slight degradation in that region (Figure~\ref{fig6}e). However, the overall velocity distribution still broadly matches the reference model throughout the entire depth range, demonstrating that the framework maintains geological plausibility even with compromised deep constraints thanks to the shallow to deep prior progression. Critically, when compared to the background-only case (Figure~\ref{fig2}b2,c2,d2,e2), the presented result with attenuated deep image shows substantially lower uncertainty (Figure~\ref{fig6}d) and error (Figure~\ref{fig6}e) in the deep section. This improvement occurs because the shallow section (above 2.5~km) benefited from high-quality structural constraints provided by the image, producing accurate shallow velocity estimates that are then progressively propagated downward using our depth-progressive synthesis. In other words, the reliable shallow velocity building, enabled by the preserved shallow structural information, guides the generation of deeper velocities even where direct structural constraints are unreliable or absent. This demonstrates a fundamental advantage of the shallow-to-deep approach: accurate VMB in well-constrained shallow regions actively improves velocity estimates at depth through explicit downward information propagation.

\subsection{Marmousi model}
\begin{figure}[htbp]
\centering
\includegraphics[width=1\textwidth]{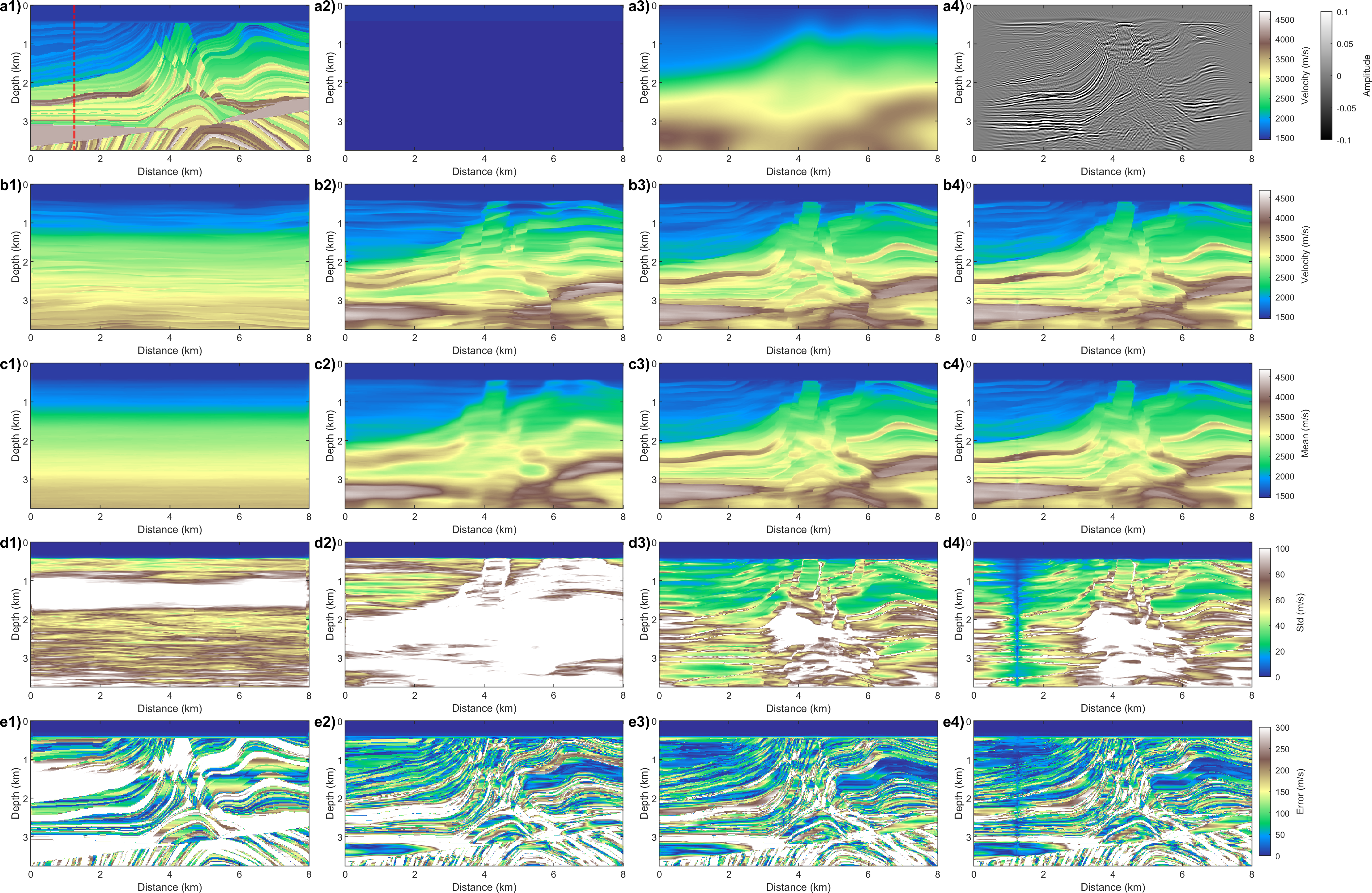}
\caption{Velocity model building results for the Marmousi model under different conditioning scenarios. First row (a1-a4): true velocity model with well location (red dashed line), shallow prior, background velocity model, and migration image. Second row (b1-b4): individual realizations generated under unconditional, background-only, background+image, and background+image+well conditioning. Third row (c1-c4): the mean of the 50 realizations for the four scenarios. Fourth row (d1-d4): the standard deviation of the 50 realizations. Fifth row (e1-e4): absolute error between the mean (row c) and true model (a1).}
\label{fig7}
\end{figure}

We, then, test our method on the complex Marmousi model. Here, the Marmousi model represents our true model, as the entire training dataset is constructed using velocity realizations guided by the three wells extracted from this model. Testing on Marmousi therefore evaluates how well our framework generalizes to the complex subsurface structure, a critical assessment for practical deployment. The Marmousi data are simulated using the same setup described in the training set, as that setup was chosen to match the field data, as we consider here the Marmousi synthetic data as our field data. The peak frequency of the Ricker wavelet used is 20 Hz. Compared to the previous in-distribution synthetic model, the Marmousi model (Figure~\ref{fig7}a1) presents significantly greater challenges due to its complex geological structure, particularly the prominent fault zone in the central portion of the model. Moreover, the background velocity model (Figure~\ref{fig7}a3) used for migration is insufficiently accurate in this complex region, resulting in a migration image (Figure~\ref{fig7}a4) with substantial artifacts. We can see that the central fault zone is poorly imaged and appears as a gap distorting the structural constraint. This represents a realistic scenario where migration image is incomplete or unreliable in poorly illuminated or structurally complex areas.

Figure~\ref{fig7} presents the VMB results under different conditioning scenarios, with the same layout as Figure~\ref{fig3}. The first row shows the true Marmousi velocity, shallow prior, background velocity, and migration image. Rows 2-5 display individual realizations, means, standard deviations, and absolute errors for the four conditioning cases. Without any conditions (first column), the ensemble mean and standard deviation (Figures~\ref{fig7}c1,d1) naturally fall within the learned prior distribution, closely matching the training statistics in Figure~\ref{fig2}, though with large errors (Figure~\ref{fig7}e1) as the model-specific structure is not captured. Adding background/migration velocity (second column) refines the large-scale trend but produces smooth results with widespread high uncertainty (Figures~\ref{fig7}c2,d2), indicating insufficient constraint on fine-scale variations. The structural constraint (third column) dramatically improves resolution and accuracy (Figures~\ref{fig7}c3,e3). However, the limitation of the imperfect migration image becomes evident: the central fault zone, which is poorly imaged due to the inaccurate background velocity, remains a region of high uncertainty (Figure~\ref{fig7}d3). This uncertainty correctly reflects the lack of reliable structural guidance in this area. Despite this limitation, the error map (Figure~\ref{fig7}e3) shows substantial overall improvement. Finally, adding well constraints (fourth column) provides effective local control, reducing both uncertainty and error near the well location at 1.25~km (Figures~\ref{fig7}d4,e4).

\begin{figure}[htbp]
\centering
\includegraphics[width=1\textwidth]{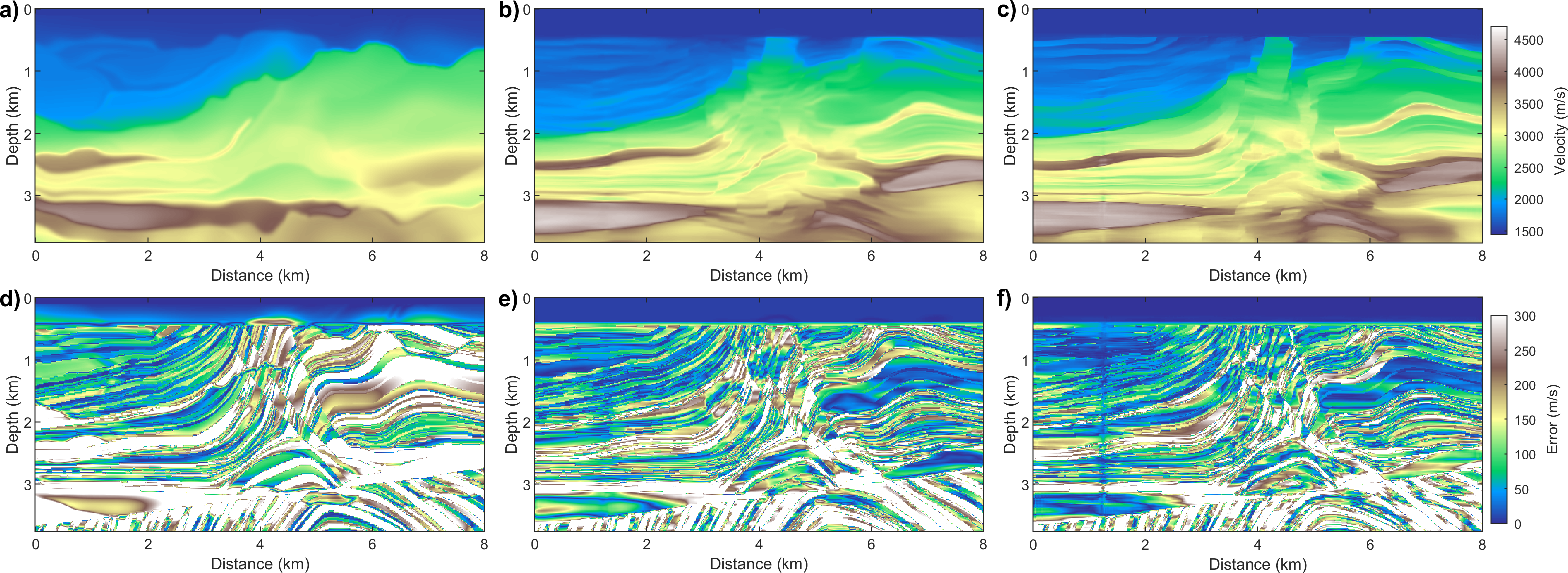}
\caption{Comparison of velocity model building results for the Marmousi model. First row (a-c): velocity models generated by WDIB, GenAIB, and our method (the mean of the 50 realizations). Second row (d-f): corresponding absolute errors relative to the true velocity model.}
\label{fig8}
\end{figure}

Comparisons with the benchmark methods is shown in Figure~\ref{fig8}. WDIB (Figure~\ref{fig8}a) struggles significantly in this scenario due to the poor quality of the migration image, which contains substantial artifacts particularly in the central fault zone. These artifacts prevent reliable dip estimation, and combined with the structural complexity and the limitation of having only a single well, WDIB fails to produce a stable velocity model and leads to a large error (Figure~\ref{fig8}d) throughout the domain. In contrast, both DL-based methods demonstrate more robust performance under these challenging conditions. GenAIB (Figure~\ref{fig8}b) achieves substantially better structural definition than WDIB, though it still exhibits noticeable errors in the complex central and deep regions (Figure~\ref{fig8}e). Our method (Figure~\ref{fig8}c) delivers the most accurate velocity model (Table \ref{tab2}), with the lowest errors throughout most of the domain (Figure~\ref{fig8}f). While the central fault zone remains challenging for all methods due to the poor migration image quality, our method handles this more gracefully by maintaining geological plausibility in the face of incomplete structural constraints.

\begin{figure}[htbp]
\centering
\includegraphics[width=1\textwidth]{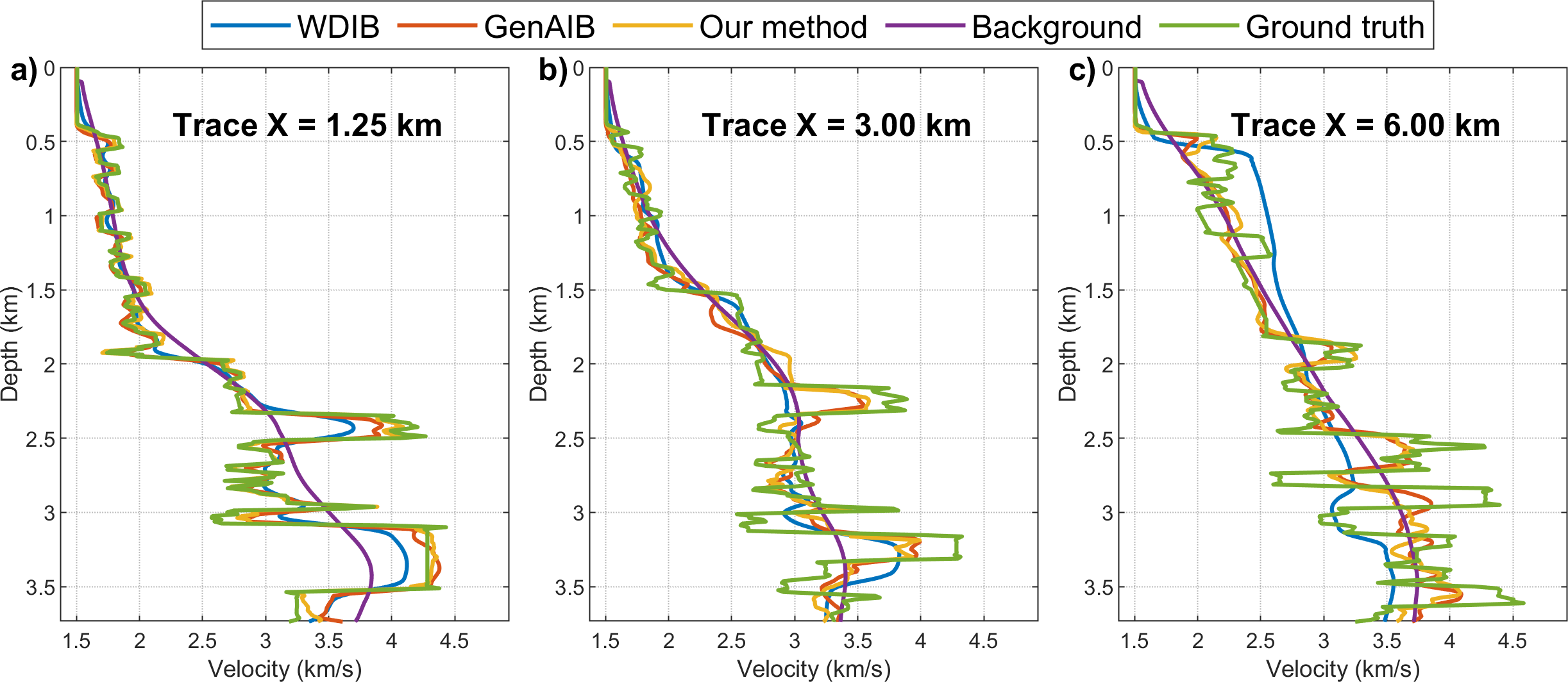}
\caption{Vertical velocity profiles for Marmousi model extracted at three lateral positions: (a) 1.25~km (well location), (b) 3.00~km, and (c) 6.00~km. The profiles compare WDIB (blue), GenAIB (orange), our method (yellow), background velocity (purple), and ground truth (green).}
\label{fig9}
\end{figure}

Figure~\ref{fig9} further presents vertical velocity profiles at 1.25~km (well location), 3.0~km, and 6.0~km. At the well location (Figure~\ref{fig9}a), our method closely tracks the true velocity throughout the profile, while WDIB over-smooths and GenAIB shows slight deviations in the deeper section. At 3.0~km (Figure~\ref{fig9}b) and 6.0~km (Figure~\ref{fig9}c), which are distant from the well, both DL-based methods demonstrate remarkably similar accuracy and closely track the ground truth, significantly outperforming WDIB. The conventional interpolation approach struggles particularly in regions far from well control, exhibiting substantial deviations from the true velocity throughout the depth profiles. 

\begin{figure}[htbp]
\centering
\includegraphics[width=1\textwidth]{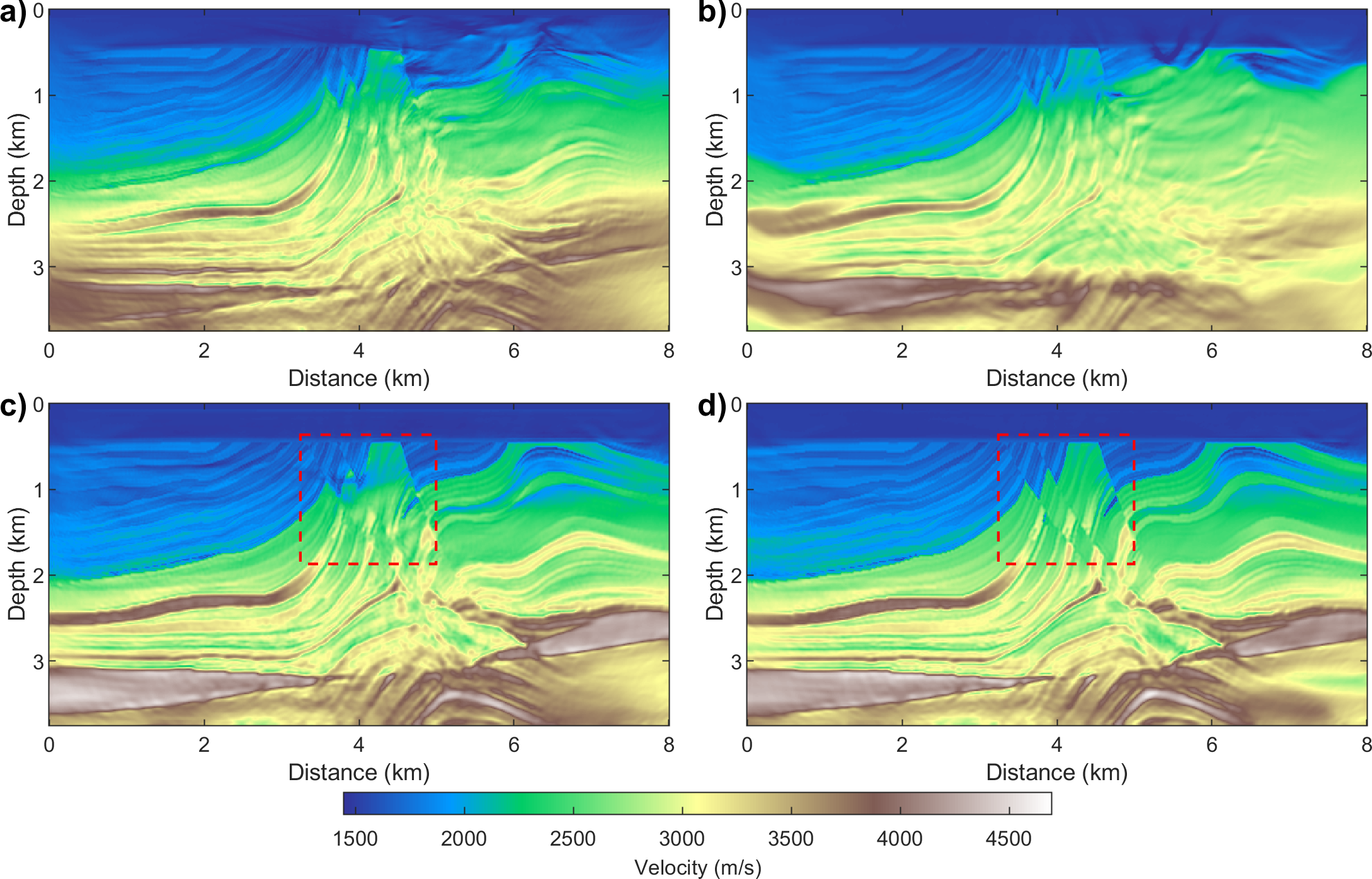}
\caption{FWI results for the Marmousi model using different initial velocity models. (a) FWI initialized from the smooth background velocity. (b-d) FWI initialized from WDIB, GenAIB, and our method, respectively. Red dashed boxes in (c) and (d) indicate the region shown in Figure~\ref{fig11}.}
\label{fig10}
\end{figure}

To further validate the quality of the velocity models constructed by different methods, we use them as starting models for FWI and compare the results against FWI initialized from the original smooth background velocity. Figure~\ref{fig10} presents the FWI results, where (a) shows the result using the background velocity as the initial velocity, and (b-d) show results using WDIB, GenAIB, and our method as initial models, respectively. Here, when we perform FWI, the frequencies below 5~Hz are removed from the observed data to simulate realistic scenarios with missing low frequencies. FWI initialized from the smooth background velocity (Figure~\ref{fig10}a) suffers from cycle skipping, which manifests as clear artifacts in the upper-right portion of the model where the inaccurate initial velocity leads to phase misalignment and convergence to a local minimum. This demonstrates the critical importance of providing FWI with a good starting model. Surprisingly, using WDIB as initialization (Figure~\ref{fig10}b) yields even worse results than the background-initialized FWI, particularly in the right portion of the model. The poor quality of the WDIB velocity model, which is caused by unreliable dip estimation from the migration image, provides misleading guidance that drives FWI further away from the true solution. This highlights the risk of conventional interpolation methods in complex scenarios where structural constraints are compromised. In stark contrast, FWI initialized from GenAIB (Figure~\ref{fig10}c) achieves substantially improved results with significantly higher resolution. The DL-based initial model successfully guides FWI to recover fine-scale layering and structural features, demonstrating the value of data-driven VMB for FWI initialization. Our method (Figure~\ref{fig10}d) produces even better FWI results, particularly evident in the central fault region. The inversion exhibits higher resolution and superior structural recovery compared to GenAIB, where the fault-bounded structures are more sharply defined and the overall geological complexity is better preserved in the final inverted model.

\begin{figure}[htbp]
\centering
\includegraphics[width=1\textwidth]{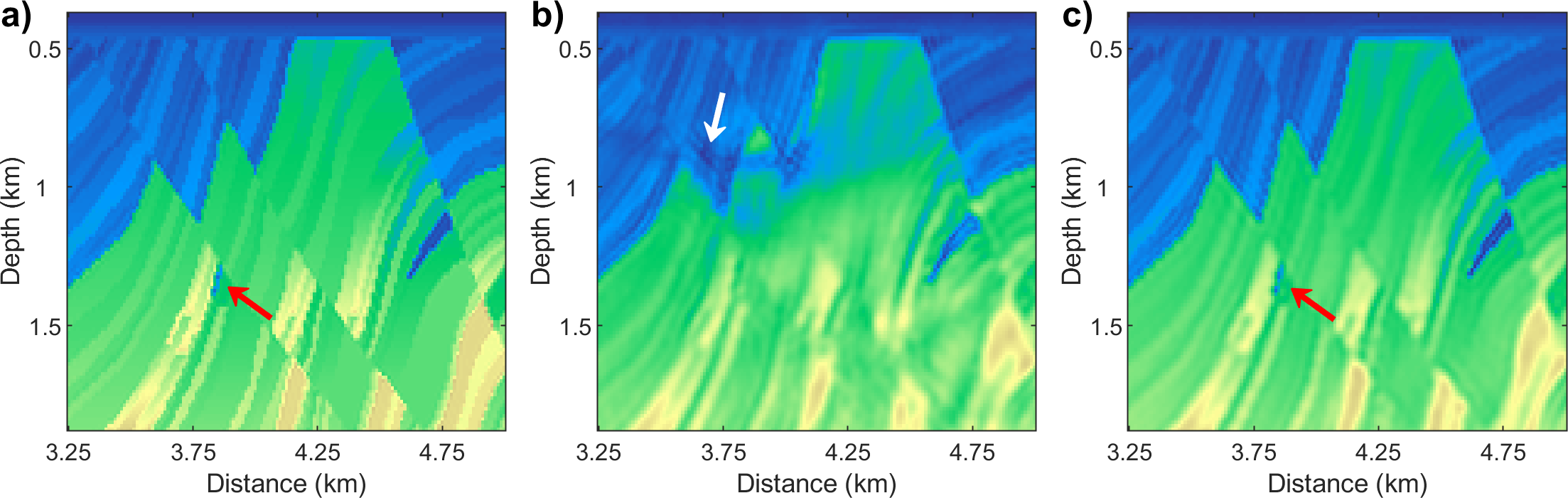}
\caption{Zoomed-in view of the central fault region from FWI results. (a) True velocity. (b) FWI result using GenAIB initialization. (c) FWI result using our method initialization.}
\label{fig11}
\end{figure}

To better compare the FWI results from GenAIB and our method, we extract and magnify the central fault region (red dashed boxes in Figures~\ref{fig10}c,d) in Figure~\ref{fig11}. The true velocity structure in this region (Figure~\ref{fig11}a) contains complex fault geometries and small-scale reservoir features. The zoomed comparison reveals that our method provides an initial model that enables FWI to achieve markedly higher resolution in recovering these complex structures. Notably, even the very small-scale reservoir body indicated by the red arrow in Figure~\ref{fig11}c is well delineated in the FWI result using our initialization. However, GenAIB-initialized FWI (Figure~\ref{fig11}b) fails to achieve comparable resolution in this challenging region. More critically, clear cycle-skipping artifacts are visible in the GenAIB result, as indicated by the white arrow. 

\begin{table}[h]
\centering
\caption{Quantitative comparison of initial velocity models and FWI results for the Marmousi model. MAE is reported in m/s, and SSIM ranges from 0 to 1, with bold text indicating best accuracy.}\label{tab2}
\begin{tabular}{l l c c c}
\toprule
 &  & WDIB & GenAIB & Our method \\
\midrule
\multirow{2}{*}{\textbf{Initial model}} 
  & MAE  & 234.08 & 155.91 & \textbf{147.33} \\
  & SSIM   & 0.582 & 0.659 & \textbf{0.681} \\
\midrule
\multirow{2}{*}{\textbf{Inverted model}} 
  & MAE  & 219.37 & 134.86 & \textbf{110.23} \\
  & SSIM  & 0.616 & 0.701 & \textbf{0.803}\\
\bottomrule
\end{tabular}
\end{table}

Quantitative assessment of the velocity models is provided in Table~\ref{tab2}, which reports MAE and SSIM metrics for both the initial velocity models and the final FWI results. For the initial models, our method achieves MAE of 147.33~m/s and SSIM of 0.681, outperforming GenAIB (155.91~m/s, 0.659) and substantially surpassing WDIB (234.08~m/s, 0.582). More importantly, the quality advantage of our initial model translates directly to superior FWI results: the inverted model using our initialization achieves MAE of 110.23~m/s and SSIM of 0.803, compared to 134.86~m/s and 0.701 for GenAIB, and 219.37~m/s and 0.616 for WDIB. These metrics confirm that our depth-progressive framework not only produces more accurate initial velocity models but also enables FWI to achieve better final inversion results. It is worth noting that the VMB accuracy metrics for the initial models are lower than those obtained in the in-distribution synthetic model, which is expected given the increased structural complexity and the challenges posed by incomplete migration imaging in the Marmousi model.

\section{\textbf{Field example}}
Having validated our framework on synthetic examples, we now test its effectiveness on field data acquired from offshore northwest Australia using variable-depth streamer, where one well is available in the survey area. The original dataset contains 1824 shot gathers with a lateral spacing of approximately 18.75~m. To reduce computational cost, we select 116 shots with a spacing of approximately 90~m for our tests.

\begin{figure}[htbp]
\centering
\includegraphics[width=1\textwidth]{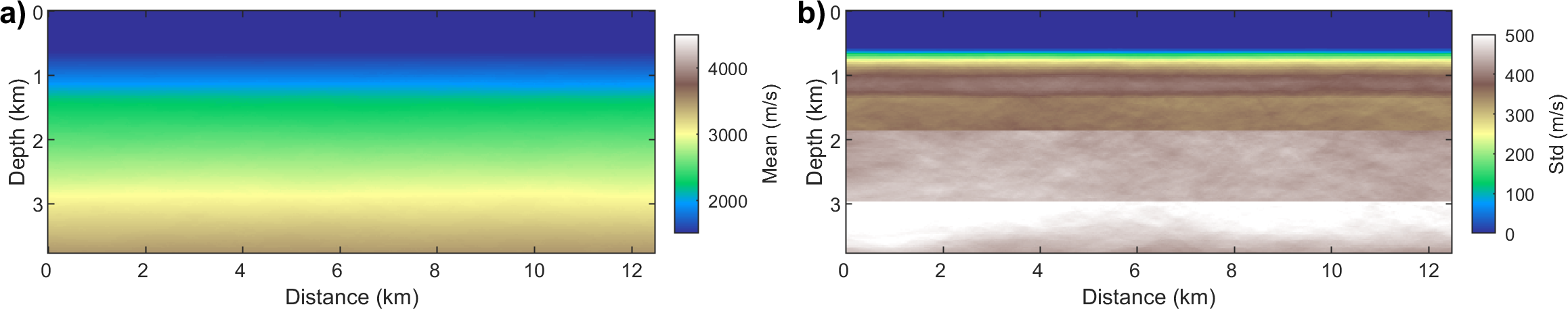}
\caption{Statistical characteristics of the training dataset for field example. (a) Mean velocity model. (b) Standard deviation map.}
\label{fig12}
\end{figure}

We construct the training dataset using the available well to guide velocity model generation, producing 1000 synthetic velocity models with dimensions of $150\times500$ grid points and a uniform grid spacing of 25~m. Figure~\ref{fig12} shows the mean and standard deviation of these models, revealing the overall velocity trend and variability embedded in the training data. For each velocity model, we extract the source wavelet from the field data and perform acoustic modeling using the same acquisition geometry as the field survey to generate synthetic shot gathers. Smooth background/migration velocity models are obtained by applying strong spatial smoothing to each true velocity model, and RTM is performed using these background models to compute the migration images. Data augmentation through horizontal flipping doubles the dataset to 2000 training samples. We adopt the same training configuration as in the synthetic examples, with training completed in approximately 23 hours on a single NVIDIA A100 GPU. For comparison, the GenAIB benchmark method is trained on the same dataset with identical settings, requiring approximately 31.5 hours.

\begin{figure}[htbp]
\centering
\includegraphics[width=1\textwidth]{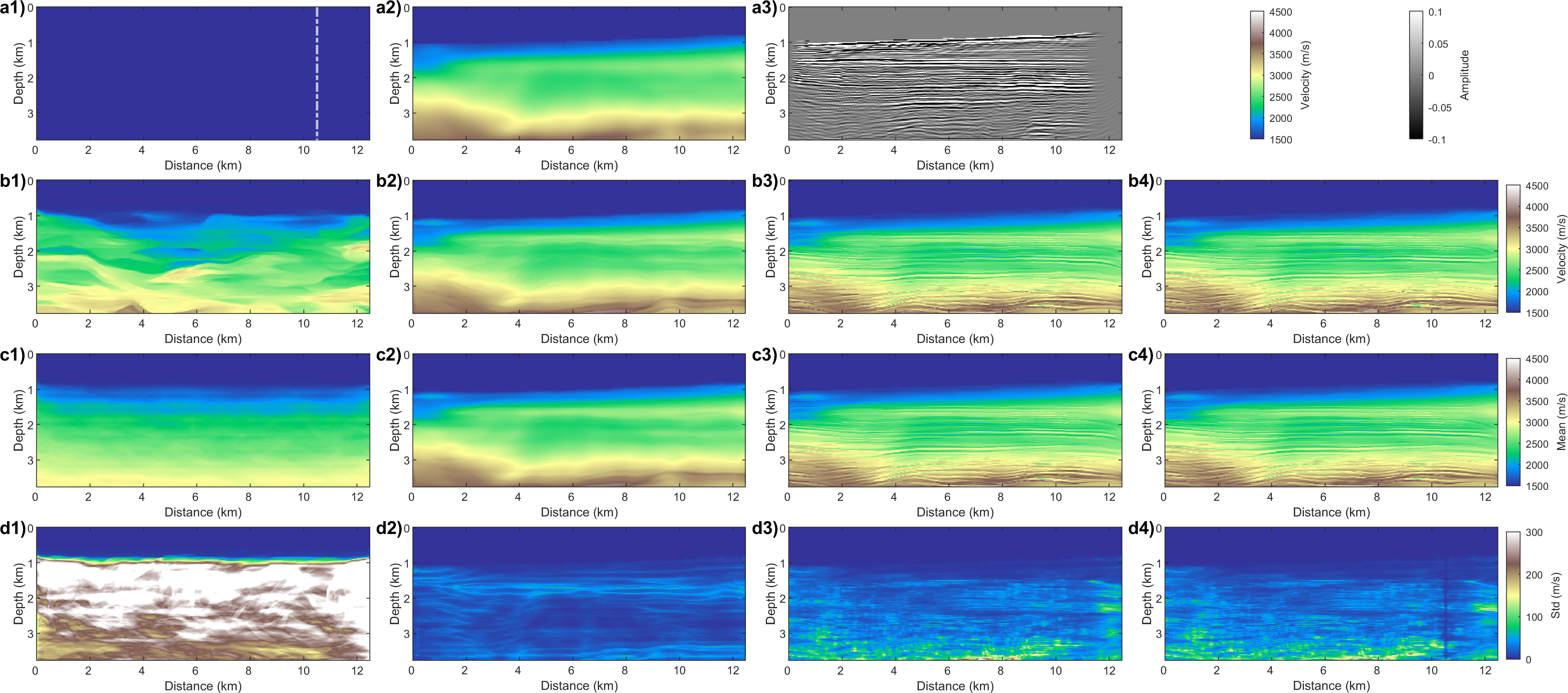}
\caption{Velocity modeling results for field data under different conditioning scenarios. First row (a1-a3): shallow prior (white dashed line denotes the well location), background velocity from migration velocity analysis, and RTM image. Second row (b1-b4): individual realizations generated under unconditional, background-only, background+image, and background+image+well conditioning. Third row (c1-c4): the mean of the 50 realizations for the four scenarios. Fourth row (d1-d4): the standard deviation of the 50 realizations.}
\label{fig13}
\end{figure}

\begin{figure}[htbp]
\centering
\includegraphics[width=1\textwidth]{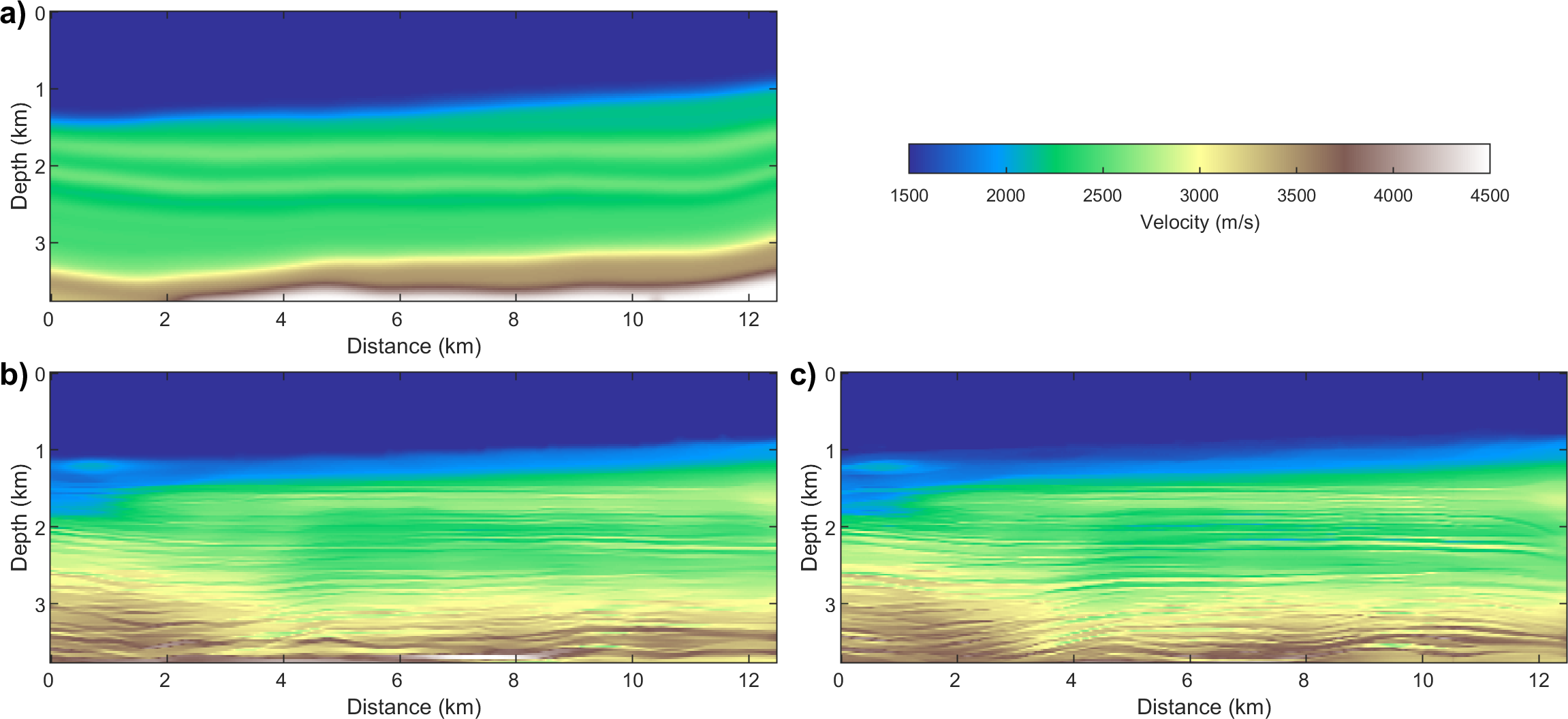}
\caption{Comparison of velocity models constructed by different methods for the field data. (a) WDIB, (b) GenAIB, and (c) our method. }
\label{fig14}
\end{figure}

After training, we test the generation performance under different conditioning scenarios on the field data, as shown in Figure~\ref{fig13}. The first row displays the shallow prior, background velocity model (obtained from migration velocity analysis), and the RTM image produced using the background velocity. Rows 2-4 show individual samples, means, and standard deviations for the four conditioning cases: unconditional, background-only, background+image, and all constraints. Under unconditional generation (first column), the samples exhibit a general velocity increase from shallow to deep, and the mean (Figure~\ref{fig13}c1) matches the training data mean (Figure~\ref{fig12}a) reasonably well. However, the standard deviation (Figure~\ref{fig13}d1) shows some differences from the training variability, with particularly high uncertainty emerging below 1~km depth, indicating substantial VMB ambiguity in the absence of specific constraints for this field scenario. Adding the background velocity constraint (second column) produces models that broadly follow the background trend while introducing slight fine-scale details. The standard deviation (Figure~\ref{fig13}d2) is significantly reduced compared to the unconditional case, confirming that the background provides effective large-scale guidance. Incorporating both background velocity and the migration image (third column) introduces additional geological details, similar to observations in the synthetic examples. However, unlike the synthetic cases, the standard deviation (Figure~\ref{fig13}d3) is slightly higher in the deep section compared to the background-only case (Figure~\ref{fig13}d2). This expected increase in uncertainty arises from propagation effects inherent in field data, where seismic attenuation and other factors cause the migrated image energy to weaken significantly at depth (visible in Figure~\ref{fig13}a3), even after depth-dependent amplitude corrections. The reduced reliability of the structural constraint at depth translates into elevated VMB uncertainty in those regions. Finally, adding well constraints (fourth column) exhibits the same behavior observed in synthetic examples: uncertainty near the well location is notably reduced (Figure~\ref{fig13}d4). 

\begin{figure}[htbp]
\centering
\includegraphics[width=0.6\textwidth]{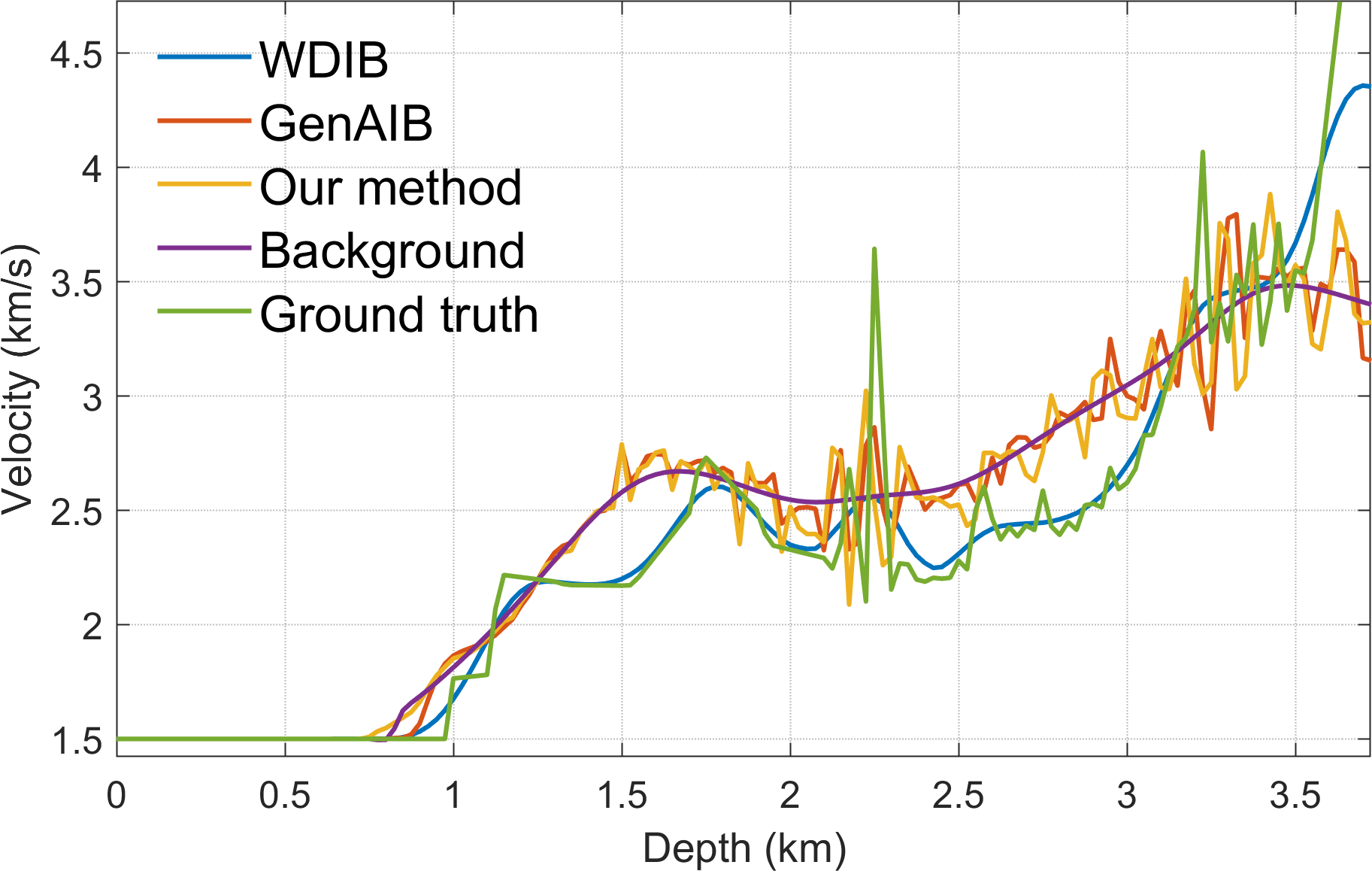}
\caption{Velocity profiles for field data extracted at the well location comparing WDIB (blue), GenAIB (orange), our method (yellow), background velocity (purple), and well log (green). }
\label{fig15}
\end{figure}

We further compare our method against WDIB and GenAIB using all available constraints, with DL-based methods generating ensemble means from 50 realizations. Figure~\ref{fig14} presents the velocity models constructed by the three methods. The WDIB result (Figure~\ref{fig14}a) is overly smooth with limited geological detail. Both DL-based methods (Figures~\ref{fig14}b,c) achieve substantially higher resolution. In which, our method (Figure~\ref{fig14}c) exhibits noticeably finer layering compared to GenAIB, revealing thin-bed structures that are absent in the GenAIB result. Figure~\ref{fig15} shows velocity profiles extracted at the well location. The WDIB profile essentially represents a smoothed version of the well velocity, lacking the fine-scale variations evident in the well log. Both DL-based methods are influenced by the background velocity constraint, following a trend closer to the background model while incorporating additional velocity variations guided by the well data and migration-derived structure. An important observation is that unlike the synthetic examples where DL-based methods closely matched the well velocities, here both methods exhibit some deviation from the well log. This arises from two factors: first, the characteristics of field migration images differ from those in the training dataset (which was based on synthetic data), leading to generalization challenges; second, well-measured velocities and seismic velocities inherently differ due to scale and measurement methodology, contributing additional mismatch.

\begin{figure}[htbp]
\centering
\includegraphics[width=1\textwidth]{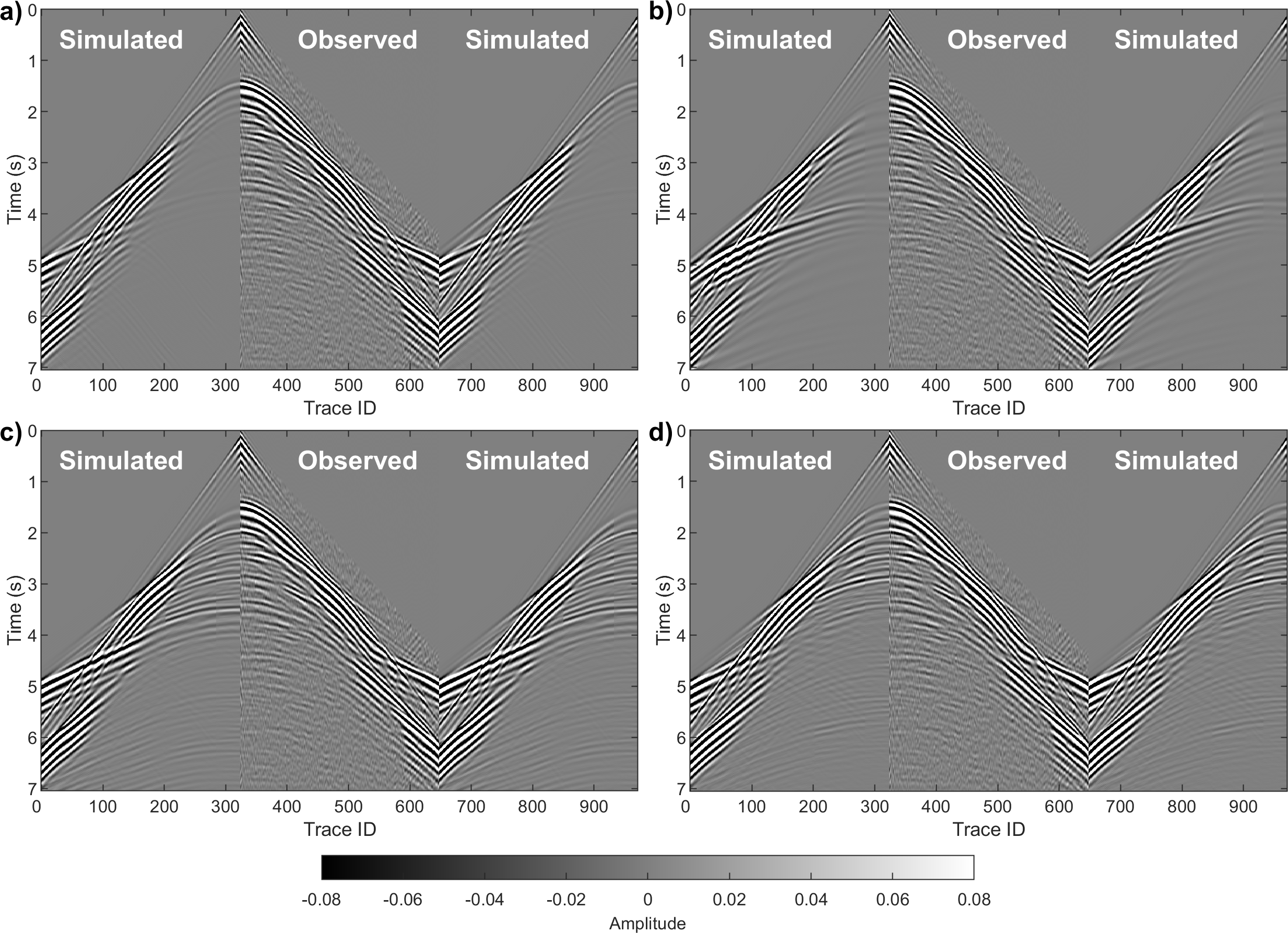}
\caption{Comparison of synthetic and observed shot gathers for a representative shot using different velocity models: (a) Background velocity, (b) WDIB, (c) GenAIB, (d) our method. Each panel shows simulated data on the left and right, with observed data in the center. }
\label{fig16}
\end{figure}

\begin{figure}[htbp]
\centering
\includegraphics[width=1\textwidth]{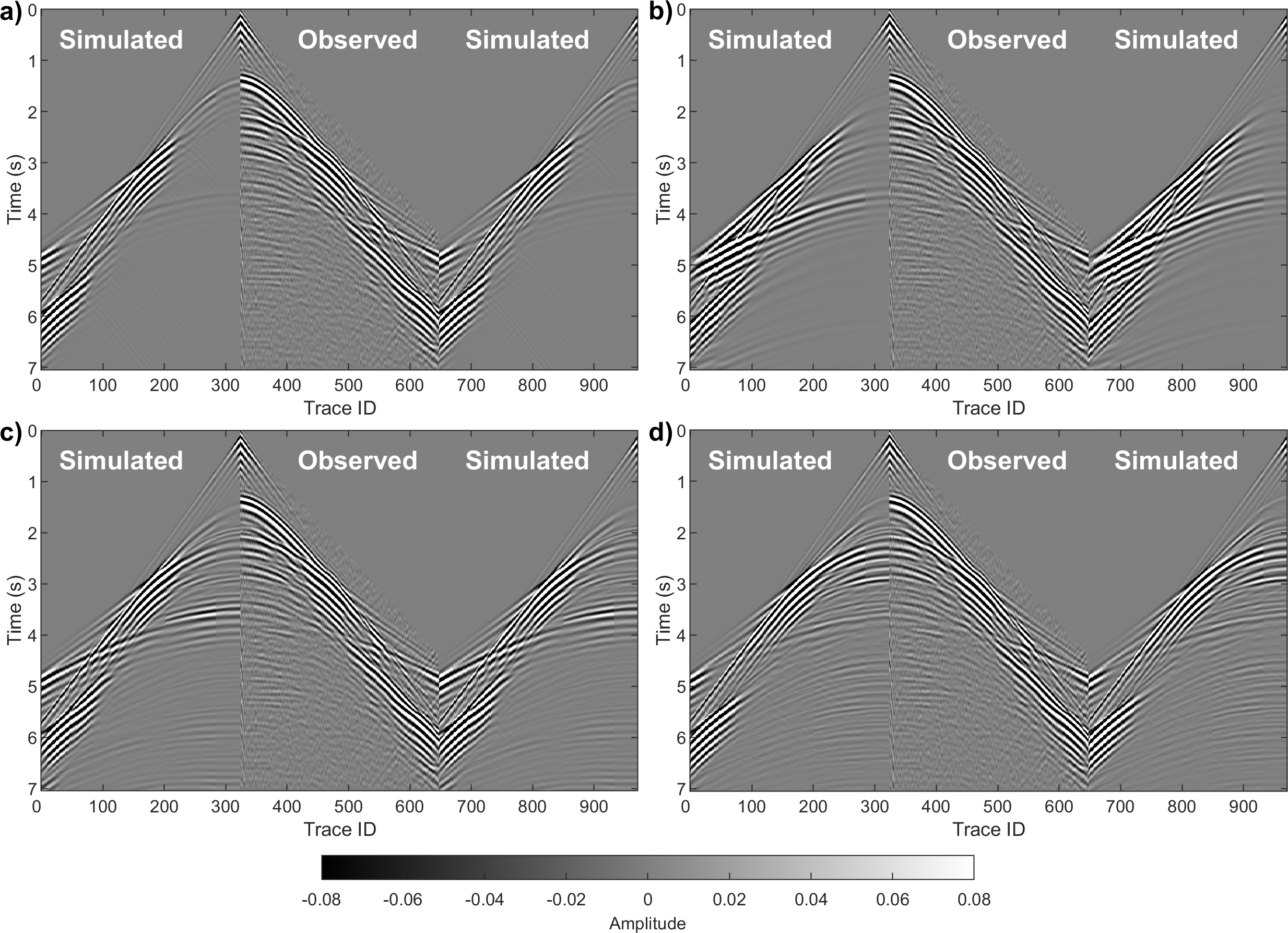}
\caption{Comparison of synthetic and observed shot gathers for a second representative shot. Layout as in Figure~\ref{fig16}. }
\label{fig17}
\end{figure}

Since no ground truth velocity model is available for field data, we cannot directly quantify VMB accuracy. Instead, we evaluate velocity model quality by comparing synthetic shot gathers generated from each model with the observed field data. We do that taking into account that unlike FWI, data fitting is not a direct objective of this approach. Figures~\ref{fig16} and~\ref{fig17} show such comparisons for two representative shot gathers, where (a) corresponds to the background velocity, and (b-d) correspond to WDIB, GenAIB, and our method, respectively. The synthetic data from both the background velocity (Figures~\ref{fig16}a and~\ref{fig17}a) and WDIB (Figures~\ref{fig16}b,~\ref{fig17}b) exhibit limited reflection events due to the smooth nature of these models. More critically, WDIB-generated data show substantial misalignment in shallow reflection events compared to the observed data. Such phase errors would drive a subsequent FWI into cycle-skipping. GenAIB (Figures~\ref{fig16}c and ~\ref{fig17}c) produces synthetic data with more reflection events due to the higher-resolution velocity model. However, significant misalignment persists across many events throughout the record, indicating that despite achieving higher resolution, the GenAIB velocity model contains substantial inaccuracies. This observation highlights a critical limitation: resolution alone does not guarantee accuracy. The velocity structure must also be kinematically correct to produce data that match observations. In contrast, our method (Figures~\ref{fig16}d and ~\ref{fig17}d) generates synthetic data that generally match the observed field data remarkably well. The reflection events align closely in both traveltime and amplitude across most of the shot gather, with only minor misalignment at far offsets. This strong data fit demonstrates that our depth-progressive framework, despite being trained on synthetic data, successfully generalizes to field conditions and produces velocity models that are both high-resolution and kinematically accurate, an essential quality for successful FWI initialization.

\section{\textbf{Discussion}}\label{sec:discussion}

Our depth-progressive diffusion framework demonstrated substantial advantages for velocity model building (VMB) in realistic exploration scenarios. By synthesizing velocity models from shallow to deep while jointly conditioning on background velocity, migration-derived structure, and sparse well constraints, the method produces high-resolution velocity models that are both geologically plausible and kinematically accurate. The synthetic and field data examples consistently show that our approach outperforms both conventional interpolation methods and alternative deep learning (DL) approaches in accuracy, structural fidelity, and uncertainty quantification. Critically, the velocity models constructed by our method provide reliable initializations for full-wabeform inversion, effectively mitigating cycle-skipping and enabling successful recovery of fine-scale subsurface features even in structurally complex areas. The probabilistic nature of the framework also provides valuable uncertainty estimates that correlate well with VMB errors, offering practitioners guidance on where additional constraints or refined processing may be needed.

\begin{table}[htbp]
\centering
\caption{Comparison of memory and time consumption between our method and GenAIB during training and inference, with bold text indicating low cost}
\label{tab3}
\renewcommand{\arraystretch}{1.25}
\begin{tabular}{lcccc}
\toprule
\multirow{2}{*}{Method} & \multicolumn{2}{c}{Training} & \multicolumn{2}{c}{Inference} \\
\cmidrule(lr){2-3} \cmidrule(lr){4-5}
& Memory (GB) & Time (s/100 iterations) & Memory (GB) & Time (s/50 samples) \\
\midrule
Our method & \textbf{28.49} & \textbf{48}  & \textbf{13.41} & 84.70 \\
GenAIB     & 72.38 & 109 & 55.26 & \textbf{21.84} \\
\bottomrule
\end{tabular}
\end{table}

Compared to recent generative diffusion model-based VMB method (denoted by GenAIB) \citep{zhang2025well}, our depth-progressive approach offers advantages in both accuracy and computational efficiency. As demonstrated in Tables~\ref{tab1} and~\ref{tab2}, our method consistently achieves higher performance across synthetic examples. Beyond accuracy, our framework exhibits significantly reduced computational cost during both training and inference. Table~\ref{tab3} compares memory and time consumption between the our method and GenAIB using identical training configurations (batch size of 8) on the Marmousi-derived synthetic dataset (600 models described in Section~\ref{sec:synthetic_example}). During training, our method requires 28.49~GB of memory and 48 seconds per 100 iterations, compared to GenAIB's 72.38~GB and 109 seconds. 
For inference, we test on the in-distribution synthetic model, generating 50 velocity realizations. Our method only consumes 13.41~GB of memory, substantially lower than GenAIB's 55.26~GB. While GenAIB completes inference faster (21.84 vs. 84.70 seconds) due to its single-pass generation, our method's reduced memory footprint enables deployment on more modest hardware and facilitates ensemble generation for uncertainty quantification. The depth-progressive strategy, by processing overlapping depth windows rather than full models, inherently limits memory requirements and enables fast training even with limited GPU resources.

Despite these advantages, our method is not without limitations. A primary challenge is the requirement to construct training datasets that reflect the characteristics of the target exploration area. This process demands careful consideration of acquisition geometry, source wavelet extraction, migration parameters, and background velocity smoothing to ensure that synthetic training data align with field data characteristics. As observed in the field data example, discrepancies between synthetic and field migration images can lead to generalization challenges, where the trained model exhibits reduced accuracy compared to in-distribution synthetic tests. A remedy to that is possibly achieved through using least square RTM as it promises more balanced amplitudes. Also, improving the model's robustness to such domain shifts requires either more diverse training data or more advanced training strategies that explicitly account for the variability of field data characteristics.

Looking forward, constructing large-scale, open-source velocity model datasets representing diverse geological settings, acquisition configurations, and data quality levels would significantly advance the practical applicability of DL-based VMB methods \citep{deng2022openfwi}. Such a dataset would enable training of foundation models with broad generalization capabilities, which could then be rapidly fine-tuned for specific exploration areas with minimal additional data. As a result, it has the potential to make advanced VMB techniques more accessible and reduce the level of specialized expertise required for their practical deployment in exploration geophysics. 
\section{\textbf{Conclusions}}

Building on the depth-progressive diffusion framework introduced in Part~I, we extended the methodology to exploration settings where perfect structural constraints are not available. By replacing idealized reflectivity constraints with migration images and incorporating smooth background velocity models as additional conditioning inputs, we adapted the framework to rely on information sources that are commonly available in practice: tomographic background/migration velocities, migrated seismic images, and sparse well measurements. The main developments include a conditioning strategy that integrates multiple imperfect constraints, a training dataset construction procedure that reflects realistic acquisition characteristics and migration artifacts, and network architecture modifications that explicitly encode background velocity information. Through mixed conditional-unconditional training, the framework can also handle cases in which some conditioning inputs are unavailable or unreliable.

Numerical tests on synthetic examples demonstrated that the proposed method yields velocity models with improved accuracy and structural consistency compared with conventional well-guided interpolation and representative deep learning baselines. The field data example further suggested that the approach is applicable to real surveys: although the network is trained on synthetic models, it produces velocity models whose simulated shot gathers are in reasonable agreement with the observed data. Overall, the ability to combine complementary information sources within a unified probabilistic framework, while providing an estimate of uncertainty, represents a step toward more data-driven velocity model building under realistic exploration conditions.

\section{Acknowledgments}
This publication is based on work supported by the King Abdullah University of Science and Technology (KAUST). The authors thank the DeepWave sponsors for their support. This work utilized the resources of the Supercomputing Laboratory at King Abdullah University of Science and Technology (KAUST) in Thuwal, Saudi Arabia.
\section{Code and data availability}

All codes, datasets, and pre-trained models associated with this work are publicly available to ensure full reproducibility of the results reported in the manuscript.

\subsection{Code availability}
The source code is openly available on GitHub at \url{https://github.com/DeepWave-KAUST/DiffVMB-pub}. The repository includes the complete implementation of the depth-progressive diffusion framework, organized into two parts corresponding to the two companion manuscripts. For Part~II, the \texttt{diffvmb\_part2/} directory contains the core Python library (\texttt{code/}), the training script (\texttt{train.py}), and the sampling/inference script (\texttt{sample.py}). Compared with Part~I, the framework is extended with two additional conditioning inputs: a migration-derived structural image obtained by applying reverse-time migration (RTM) with a smooth background velocity model, and the background velocity model itself as a low-wavenumber constraint. Installation instructions and dependencies are provided in the repository README.

\subsection{Dataset availability}
The training and test datasets are publicly available on Zenodo \citep[DOI:][]{zenodo2025diffvmb} as part of the archive \texttt{dataset.zip}. After extraction, the Part~II data are located under \texttt{dataset/part2/}, which contains two sub-directories:
\begin{itemize}
    \item \texttt{train/}: training samples in NPZ format, each containing two arrays, namely \texttt{vp} (P-wave velocity, shape $n_z \times n_x$) and \texttt{mig} (migration-derived structural image, same shape), which represents 2-D sections extracted from industrial velocity models;
    \item \texttt{test/}: two benchmark velocity models in MAT format used for evaluation, comprising one in-distribution model (Syn) and one out-of-distribution model (Marmousi) to assess generalization capability.
\end{itemize}

\subsection{Pre-trained model}
The pre-trained model weights for Part~II are available on the same Zenodo record \citep[DOI:][]{zenodo2025diffvmb} as part of the archive \texttt{trained\_model.zip}. After extraction, the weights are stored in \texttt{trained\_model/model\_part2.pt}. The model was trained on a single NVIDIA A100 GPU. To reproduce the numerical results reported in this manuscript, users may download the pre-trained weights and the test dataset, place them in the directories specified in the repository README, and execute:
\begin{verbatim}
cd diffvmb_part2
python sample.py
\end{verbatim}

\bibliographystyle{unsrtnat}
\bibliography{references}

\end{document}